\documentclass[iop]{emulateapj}
\usepackage{aas_macros}

\usepackage{color}
\usepackage[dvipsnames,svgnames]{xcolor}

\usepackage{natbib}

\usepackage{amsmath}
\usepackage{xspace}
\usepackage{lineno}

\mathchardef\mhyphen="2D

\newlength{\dhatheight}

\newcommand{\code}[1]{\texttt{#1}\xspace}

\providecommand\physrep{\ref@jnl{Phys.~Rep.}}%
\providecommand\apjs{\ref@jnl{ApJS}}%
\providecommand{\jcap}{\ref@jnl{JCAP}}%

\newcommand{\lsim}{\lower0.6ex\vbox{\hbox{$ \buildrel{\textstyle <}\over{\sim}\ $}}}

\shorttitle{Fornax dwarf spheroidal galaxy in the DES data}
\shortauthors{Wang et~al.}
\begin{document}

\title{The morphology and structure of stellar populations in the Fornax dwarf spheroidal galaxy from Dark Energy Survey Data}


\def\andname{}

\author{
M.~Y.~Wang\altaffilmark{1,2},
T.~de Boer\altaffilmark{3,4},
A.~Pieres\altaffilmark{5,6},
T.~S.~Li\altaffilmark{7,8},
A.~Drlica-Wagner\altaffilmark{7},
S.~E.~Koposov\altaffilmark{1,4},
A.~K.~Vivas\altaffilmark{9},
A. B.~Pace\altaffilmark{2},
B.~Santiago\altaffilmark{5,6},
A.~R.~Walker\altaffilmark{9},
D.~L.~Tucker\altaffilmark{7},
L.~Strigari\altaffilmark{2},
J.~L.~Marshall\altaffilmark{2},
B.~Yanny\altaffilmark{7},
D.~L.~DePoy\altaffilmark{2},
K.~Bechtol\altaffilmark{10},
A.~Roodman\altaffilmark{11,12},
T.~M.~C.~Abbott\altaffilmark{9},
F.~B.~Abdalla\altaffilmark{13,14},
S.~Allam\altaffilmark{7},
J.~Annis\altaffilmark{7},
S.~Avila\altaffilmark{15},
E.~Bertin\altaffilmark{16,17},
D.~Brooks\altaffilmark{13},
D.~L.~Burke\altaffilmark{11,12},
A.~Carnero~Rosell\altaffilmark{6,18},
M.~Carrasco~Kind\altaffilmark{19,20},
C.~E.~Cunha\altaffilmark{11},
C.~B.~D'Andrea\altaffilmark{21},
L.~N.~da Costa\altaffilmark{6,18},
J.~De~Vicente\altaffilmark{22},
S.~Desai\altaffilmark{23},
T.~F.~Eifler\altaffilmark{24,25},
J.~Estrada\altaffilmark{7},
B.~Flaugher\altaffilmark{7},
J.~Frieman\altaffilmark{7,8},
J.~Garc\'ia-Bellido\altaffilmark{26},
D.~W.~Gerdes\altaffilmark{27,28},
D.~Gruen\altaffilmark{11,12},
R.~A.~Gruendl\altaffilmark{19,20},
G.~Gutierrez\altaffilmark{7},
D.~L.~Hollowood\altaffilmark{29},
K.~Honscheid\altaffilmark{30,31},
D.~J.~James\altaffilmark{32},
K.~Kuehn\altaffilmark{33},
N.~Kuropatkin\altaffilmark{7},
O.~Lahav\altaffilmark{13},
M.~A.~G.~Maia\altaffilmark{6,18},
R.~Miquel\altaffilmark{34,35},
E.~Sanchez\altaffilmark{22},
V.~Scarpine\altaffilmark{7},
I.~Sevilla-Noarbe\altaffilmark{22},
M.~Smith\altaffilmark{36},
R.~C.~Smith\altaffilmark{9},
F.~Sobreira\altaffilmark{37,6},
E.~Suchyta\altaffilmark{38},
M.~E.~C.~Swanson\altaffilmark{20},
G.~Tarle\altaffilmark{28}
\\ \vspace{0.2cm} (DES Collaboration) \\
}

\affil{$^{1}$ Department of Physics, Carnegie Mellon University, Pittsburgh, PA 15213}
\affil{$^{2}$ George P. and Cynthia Woods Mitchell Institute for Fundamental Physics and Astronomy, and Department of Physics and Astronomy, Texas A\&M University, College Station, TX 77843, USA}
\affil{$^{3}$ Department of Physics, University of Surrey, Guildford GU2 7XH, UK}
\affil{$^{4}$ Institute of Astronomy, University of Cambridge, Madingley Road, Cambridge CB3 0HA, UK}
\affil{$^{5}$ Instituto de F\'\i sica, UFRGS, Caixa Postal 15051, Porto Alegre, RS - 91501-970, Brazil}
\affil{$^{6}$ Laborat\'orio Interinstitucional de e-Astronomia - LIneA, Rua Gal. Jos\'e Cristino 77, Rio de Janeiro, RJ - 20921-400, Brazil}
\affil{$^{7}$ Fermi National Accelerator Laboratory, P. O. Box 500, Batavia, IL 60510, USA}
\affil{$^{8}$ Kavli Institute for Cosmological Physics, University of Chicago, Chicago, IL 60637, USA}
\affil{$^{9}$ Cerro Tololo Inter-American Observatory, National Optical Astronomy Observatory, Casilla 603, La Serena, Chile}
\affil{$^{10}$ LSST, 933 North Cherry Avenue, Tucson, AZ 85721, USA}
\affil{$^{11}$ Kavli Institute for Particle Astrophysics \& Cosmology, P. O. Box 2450, Stanford University, Stanford, CA 94305, USA}
\affil{$^{12}$ SLAC National Accelerator Laboratory, Menlo Park, CA 94025, USA}
\affil{$^{13}$ Department of Physics \& Astronomy, University College London, Gower Street, London, WC1E 6BT, UK}
\affil{$^{14}$ Department of Physics and Electronics, Rhodes University, PO Box 94, Grahamstown, 6140, South Africa}
\affil{$^{15}$ Institute of Cosmology \& Gravitation, University of Portsmouth, Portsmouth, PO1 3FX, UK}
\affil{$^{16}$ CNRS, UMR 7095, Institut d'Astrophysique de Paris, F-75014, Paris, France}
\affil{$^{17}$ Sorbonne Universit\'es, UPMC Univ Paris 06, UMR 7095, Institut d'Astrophysique de Paris, F-75014, Paris, France}

\affil{$^{18}$ Observat\'orio Nacional, Rua Gal. Jos\'e Cristino 77, Rio de Janeiro, RJ - 20921-400, Brazil}
\affil{$^{19}$ Department of Astronomy, University of Illinois at Urbana-Champaign, 1002 W. Green Street, Urbana, IL 61801, USA}
\affil{$^{20}$ National Center for Supercomputing Applications, 1205 West Clark St., Urbana, IL 61801, USA}
\affil{$^{21}$ Department of Physics and Astronomy, University of Pennsylvania, Philadelphia, PA 19104, USA}
\affil{$^{22}$ Centro de Investigaciones Energ\'eticas, Medioambientales y Tecnol\'ogicas (CIEMAT), Madrid, Spain}
\affil{$^{23}$ Department of Physics, IIT Hyderabad, Kandi, Telangana 502285, India}
\affil{$^{24}$ Department of Astronomy/Steward Observatory, 933 North Cherry Avenue, Tucson, AZ 85721-0065, USA}
\affil{$^{25}$ Jet Propulsion Laboratory, California Institute of Technology, 4800 Oak Grove Dr., Pasadena, CA 91109, USA}
\affil{$^{26}$ Instituto de Fisica Teorica UAM/CSIC, Universidad Autonoma de Madrid, 28049 Madrid, Spain}
\affil{$^{27}$ Department of Astronomy, University of Michigan, Ann Arbor, MI 48109, USA}
\affil{$^{28}$ Department of Physics, University of Michigan, Ann Arbor, MI 48109, USA}
\affil{$^{29}$ Santa Cruz Institute for Particle Physics, Santa Cruz, CA 95064, USA}
\affil{$^{30}$ Center for Cosmology and Astro-Particle Physics, The Ohio State University, Columbus, OH 43210, USA}
\affil{$^{31}$ Department of Physics, The Ohio State University, Columbus, OH 43210, USA}
\affil{$^{32}$ Harvard-Smithsonian Center for Astrophysics, Cambridge, MA 02138, USA}
\affil{$^{33}$ Australian Astronomical Observatory, North Ryde, NSW 2113, Australia}
\affil{$^{34}$ Instituci\'o Catalana de Recerca i Estudis Avan\c{c}ats, E-08010 Barcelona, Spain}
\affil{$^{35}$ Institut de F\'{\i}sica d'Altes Energies (IFAE), The Barcelona Institute of Science and Technology, Campus UAB, 08193 Bellaterra (Barcelona) Spain}
\affil{$^{36}$ School of Physics and Astronomy, University of Southampton,  Southampton, SO17 1BJ, UK}
\affil{$^{37}$ Instituto de F\'isica Gleb Wataghin, Universidade Estadual de Campinas, 13083-859, Campinas, SP, Brazil}
\affil{$^{38}$ Computer Science and Mathematics Division, Oak Ridge National Laboratory, Oak Ridge, TN 37831}

\begin{abstract}
Using deep wide-field photometry three-year data (Y3) from the Dark Energy Survey (DES), we present a panoramic study of the Fornax dwarf spheroidal galaxy. The data presented here -- a small subset of the full survey -- uniformly covers a region of 25 square degrees centered on the galaxy to a depth of \textit{g}$\sim$ 23.5. We use this data to study the structural properties of Fornax, overall stellar population, and its member stars in different evolutionary phases. We also search for possible signs of tidal disturbance. Fornax is found to be significantly more spatially extended than what early studies suggested. No statistically significant distortions or signs of tidal disturbances were found down to a surface brightness limit of $\sim$ 32.1 mag/${\rm arcsec^{2}}$. However, there are hints of shell-like features located $\sim 30{\mathrm '}-40{\mathrm '}$ from the center of Fornax that may be stellar debris from past merger events. We also find that intermediate age and young main-sequence populations show different orientation at the galaxy center and have many substructures. The deep DES Y3 data allows us to characterize the age of those substructures with great accuracy, both those previously known and those newly discovered in this work, on the basis of their color-magnitude diagram morphology. We find that the youngest overdensities are all found on the Eastern side of Fornax, where the Fornax field population itself is slightly younger than in the West. The high quality DES Y3 data reveals that Fornax has many rich structures, and provides insights into its complex formation history. 
\end{abstract}


\section{INTRODUCTION}
\label{intro}
According to the standard $\Lambda$CDM cosmology, the Universe as we see it today is built via a scale-free process in which matter collapses under gravity and iteratively forms larger and larger structures. Dark matter dominates this hierarchical structure formation process. Dwarf galaxies, the earliest formed structures and the most dark matter dominated objects known, are therefore crucial laboratories for studying galaxy formation and evolution processes~\citep[e.g.][]{Kauffmann_etal93}. Considerable works have been invested in the study of dwarf galaxies in the Local Group (LG), as summarized recently in \cite{McConnachie_12}.

Even though they have small physical sizes, the proximity of these galaxies results in large angular extents on the sky, of approximately a few degrees. Hence the study of these systems has particularly flourished since the advent of wide-area imagers and multi-object spectrographs, which has allowed us to determine the properties of their resolved stellar component in great detail out to their very low surface brightness outskirts. This applies in particular to the ``brightest" of the Milky Way (MW) dwarf spheroidal (dSph) galaxies, those with luminosities $-$15$\lsim M_V \lsim$$-$8, for which one can gather wide-area photometric data-sets with large statistics to reach well below the oldest main-sequence turn-off (MSTO), and spectroscopic data-sets to provide velocities and metallicities of several hundreds of stars per galaxy. It is only through a joint analysis of the entire stellar sample that one can obtain a complete picture of the morphology and populations of dwarf galaxies. 

The Fornax dSph is one of the more intensively studied dSphs of the Milky Way. First discovered by \cite{Shapley_etal38}, it is one of the most luminous LG dwarf galaxies, second only to the Sagittarius dSph in terms of luminosity ($M_V$ =-13.4 $\pm$ 0.3) amongst the MW classical dSphs. Fornax has a total~(dynamical) mass of 1.6$\times$10$^{8}$ M$_{\odot}$ at a distance of~138$\pm$8 kpc~((m-M)$_{\mathrm{V}}$=20.84$\pm$0.04)~\citep{Walker_etal06, Rizzi_etal07,Pietrzynski_etal09, Lokas_etal09}. Fornax is one of the few LG dwarf galaxies that has experienced a composite, extended star formation history covering ancient ($>$10 Gyr), intermediate age (2$-$8 Gyr) and extremely young (100 Myr) stars~\citep[e.g.,][]{Gallart_etal05, de_Boer_etal2012, del_Pino_etal2013}. Spectroscopic studies of individual stars have found Fornax to have a broad metallicity distribution with a dominant metal-rich~([Fe/H]$\approx$$-$0.9 dex) component~\citep[e.g.,][]{Pont_etal04, Battaglia_etal06, Letarte_etal10}. Studies of the kinematics have further shown the presence of at least three kinematically distinct populations~\citep{Battaglia_etal06, Amorisco_etal2012}, indicating it has undergone a complex formation process. Finally, Fornax also contains several globular clusters~\citep{Shapley_etal38, Hodge_etal61A}, some of which are among the most metal-poor clusters found in the LG~\citep{Letarte_etal06, deBoer_etal13} 

Fornax can be easily resolved into individual stars down to the base of the red giant branch (RGB) with relatively short exposures on 4m-class telescopes. However, given its substantial size on the sky~(r$_{\mathrm{tidal}}$=1.18$\pm$0.07 degrees or 2.85$\pm$0.16 kpc), a detailed study of its stellar content requires wide-field imaging capabilities. Wide-field studies of Fornax made with photographic plates by~\citet{Hodge_etal61B} revealed that the ellipticity of Fornax isophotes increases with radial distance. The stellar density has also been shown to be asymmetric, with a peak density offset from the central position~\citep{deVaucouleurs_etal68, Hodge_etal74}. Resolved stellar studies of the stellar content within the tidal radius have shown that Fornax contains a radial population gradient, with younger, more metal-rich stars found preferentially towards the center~\citep{Battaglia_etal06, Coleman_etal08, de_Boer_etal2012}.

Substructure studies of Fornax have found a number of significant overdensities, which provide further signs of its complex formation scenario and possible accretion of (enriched) gas or sub-halos. Two young ($<$2 Gyr) overdensities were found by~\citet{Coleman_etal04,Coleman_etal05b}, one of which was later found to be due to background galaxies \citep{Bate_etal15}. A further three overdensities were found more towards the center of Fornax, all containing young stellar populations and metallicities (as derived from their color-magnitude diagrams (CMDs)) similar to the young Fornax field stars~\citep{deBoer_etal13, Bate_etal15}. This may indicate a connection between the over-dense components and the field stars~\citep{Olszewski_etal06}, providing tentative clues to the formation of Fornax, possibly through re-accretion of previously expelled gas~\citep[e.g.,][]{Pasetto_etal11, Nichols_etal12, Yozin_etal12}.

Despite the copious amount of work done in classifying the stellar content of Fornax, a detailed study of its extra-tidal structure has been lacking due to the need for deep ($g\approx$24), wide-field and homogeneous data in multiple filters, necessary to constrain the distribution of very low surface brightness features in the galaxy outskirts. A study covering the outskirts of Fornax was conducted recently using data from the VST/ATLAS survey~\citep{Bate_etal15}, which found no evidence of strong extra-tidal features. However, their data was only marginally able to reach the MSTO of Fornax, resulting in only a low number of tracer stars beyond the tidal radius. In contrast, the Fornax system and its environs have been covered recently by the DES survey using deep and homogeneous imaging from the Dark Energy Camera (DECam; \citealt{Flaugher_etal15}) mounted on the 4-m Blanco Telescope at Cerro Tololo Inter-American Observatory in Chile \citep{Flaugher_etal15}. The resulting data is deeper than any previous wide-field studies such as \citet{Bate_etal15} and therefore allows for a better selection of different stellar tracer features from the CMD to reduce contamination in its low luminosity outskirts.

In this article we present an extensive wide-field study of Fornax's stellar population using deep \textit{g} and \textit{r} band photometry reaching down to the MSTO stars from DES Y3 data that covers 25 ${\rm deg^2}$ of the Fornax dSph. We perform a detailed and quantitative analysis of the structural properties of this galaxy. In particular, we investigate the extra-tidal stellar content of Fornax to probe for the existence of tidal features. Given its brightness and stellar kinematics, Fornax is one of the most likely galaxies in the LG to be affected by MW tidal forces and likely to contain distorted outer isophotes and possible tidal tails \citep{Wang_etal17}.

The article is organized as follows. In Section~\S~\ref{observations} we will describe the specifics of the DES survey and the reduction of the data. Section~\S~\ref{sec:profiles} will discuss the structural parameters of the Fornax dwarf galaxy, followed by a discussion of the features in the CMD in Section~\S~\ref{sec:CMD}. The surface density maps and search for tidal features is described in Section~\S~\ref{sec:MF} and the spatial distribution of different stellar populations is studied in Section~\S~\ref{sec:sub distribution}. Finally, Section~\S~\ref{sec:overdensity} discusses the search for stellar substructures, followed by the summary and conclusions in Section~\S~\ref{conclusions}.

\section{DATA}
\label{observations}
DES is a deep, wide-area imaging survey in the ${\it grizY}$ bands performed with the DECam mounted on the 4-m Blanco Telescope at Cerro Tololo Inter-American Observatory in Chile. Here, we use wide-field imaging data from the first three years of DES operation (DES Y3). DES Y3 is the first DES data set to cover the full DES footprint, which covers $\sim$ 5000 $deg^2$ of the southern Galactic cap including the Fornax dSph galaxy area.

The details of the DES Y3 data reduction can be found in \cite{Morganson_etal18}, and the definition of the star-galaxy classification model can be found in \cite{Shipp_etal18}.
Here we briefly describe some of the procedures. The DES Y3 images were processed by the DES data management pipeline.  Individual exposures were remapped to a consistent pixel grid and coadded to increase imaging depth \citep{Morganson_etal18}.  Object detection was performed on a combination of the \textit{r}+\textit{i}+\textit{g} coadded images using the \code{SExtractor} toolkit \citep{Bertin_etal96,Bertin_etal02} with an object detection threshold of $\sim$5$\sigma$ \citep{Morganson_etal18}. The \code{ngmix} method \citep{Sheldon_14}, which fits the flux and morphology of each source over all individual single-epoch images simultaneously, was applied to overcome the difficulties of performing precise photometric and morphological measurements in the coadds.

To select a high-quality stellar sample, the \code{ngmix} method was used to fit a composite galaxy model (bulge plus disk) to each source in all bands simultaneously \citep{Drlica-Wagner_etal2018}. The best-fit size, \code{CM\_T}, and associated uncertainty, \code{CM\_T\_ERR}, from this galaxy-model fit were then used to distinguish point-like objects from those that are spatially extended. Specifically, we defined an extended classification variable, \code{NGMIX\_CLASS}, based on the sum of three selection criteria, 

\begin{eqnarray}
\code{NGMIX\_CLASS} =  & ((\code{CM\_T}+5 \; \code{CM\_T\_ERR}) > 0.1)  \nonumber \\
+ & ((\code{CM\_T}+\code{CM\_T\_ERR}) > 0.05) \nonumber \\
+ & ((\code{CM\_T}-\code{CM\_T\_ERR}) > 0.02).  
\end{eqnarray}

Our final stellar sample used \code{NGMIX\_CLASS} $\leq$ 1 criteria. This stellar selection was designed to yield a compromise between completeness and purity in the resulting stellar sample. It was compared with deeper imaging data from Hyper Suprime Cam DR1 \citep{Aihara_etal18} (see Figure 1 in \citealt{Shipp_etal18}). It is found that, in uncrowded regions, this selection is $>$ 90$\%$ complete for \textit{g} = 23.5 with a galaxy contamination rising from $\lsim$ 5$\%$ at \textit{g} $\leq$ 22.5 to $\sim$30$\%$ by \textit{g} $\sim$ 23.5. We constrained our stellar sample to the range 19 $<$ \textit{g}, \textit{r} $<$ 23.5. The faint end limit was imposed to avoid spurious density fluctuations resulting from inhomogeneous survey depth and galaxy contamination.  

\begin{table*}
\caption{Fornax structural parameters derived with the MCMC method for different profile models. The last two columns refers to the structural parameters derived by \cite{Bate_etal15} (Bate15) and \cite{Battaglia_etal06} (B06). }
{\renewcommand{\arraystretch}{1.8}
\renewcommand{\tabcolsep}{0.25cm}
\centering
\begin{tabular}{ l c c c c c c}
\hline 
\hline
Parameter & King & Exponential & Plummer & Sersic   & Bate15  & B06  \\
\hline
$\alpha_{2000}$ & ${\rm 2^h39^m53^s}$  & ${\rm 2^h39^m53^s}$ &${\rm 2^h39^m53^s}$ &${\rm 2^h39^m53^s}$& ${\rm 2^h39^m52^s}$& ${\rm 2^h39^m52^s}$\\
$\delta_{2000}$& ${\rm -34^{\circ} 30{\mathrm '}32{\mathrm ''}}$ & ${\rm -34^{\circ} 30{\mathrm '}21{\mathrm ''}}$ & ${\rm -34^{\circ} 30{\mathrm '}25{\mathrm ''}}$ & ${\rm -34^{\circ} 30{\mathrm '}32{\mathrm ''}}$& ${\rm -34^{\circ} 30{\mathrm '}39{\mathrm ''}}$ & ${\rm -34^{\circ} 30{\mathrm '}49{\mathrm ''}}$\\
Ellipticity& 0.31 $\pm$ 0.002 & 0.31 $\pm$ 0.002 & 0.29 $\pm$ 0.002 & 0.31 $\pm$ 0.002 &  0.31 $\pm$ 0.01& 0.30 $\pm$ 0.01\\
Position angle & $42.2^{\circ}$ $\pm$ $0.2^{\circ}$ & $43.2^{\circ}$ $\pm$ $0.2^{\circ}$ & $42.7^{\circ}$ $\pm$ $0.3^{\circ}$& $42.3^{\circ}$ $\pm$ $0.2^{\circ}$&  $41.6^{\circ}$ $\pm$ $0.2^{\circ}$& $46.8^{\circ}$ $\pm$ $1.6^{\circ}$\\
Sersic $r_s$ &-- & -- & -- & $16.4{\mathrm '} \pm 0.2{\mathrm '}$&  $14.5{\mathrm '} \pm 0.1{\mathrm '}$& $17.3{\mathrm '} \pm 0.2{\mathrm '}$ \\
Sersic index m&--  &--  &--  & 0.80 $\pm$ 0.006 & 0.78 $\pm$ 0.005& 0.71 $\pm$ 0.01\\
Exponential $r_e$ & --& $11.8{\mathrm '} \pm 0.04{\mathrm '}$  & -- & -- & -- & 11.0${\mathrm '} \pm 0.1{\mathrm '}$\\
Plummer $r_p$ & -- & -- & $19.9{\mathrm '} \pm 0.06{\mathrm '}$  & -- & -- & 19.6${\mathrm '} \pm 0.1{\mathrm '}$\\
King $r_t$ & $77.5{\mathrm '} \pm 0.4{\mathrm '}$  &--  &--  &-- &  $69.7{\mathrm '} \pm 0.3{\mathrm '}$ &$69.1{\mathrm '} \pm 0.4{\mathrm '}$ \\
King $r_c$ & $20.3{\mathrm '} \pm 0.1{\mathrm '}$ & -- & -- &-- &  $14.6{\mathrm '} \pm 0.1{\mathrm '}$ & $17.3{\mathrm '} \pm 0.2{\mathrm '}$ \\
\vspace{-0.5em}
2D Half-light $r_h$&20.8${\mathrm '}$ & 19.8${\mathrm '}$ & $19.4{\mathrm '}$&$20.1{\mathrm '}$ &$17.0{\mathrm '}$ (King) &$18.1{\mathrm '}$ (King)\\
\vspace{-0.5em}
 & & & & &$17.2{\mathrm '}$ (Sersic) & $18.6{\mathrm '}$ (Sersic) \\
 & & & & & & $18.5{\mathrm '}$ (Exp.) \\
$\chi^2_{reduced}$ &1.89 & 2.04 & 2.54 & 1.41 \\
\vspace{-1.5em} \\
\hline
\end{tabular}\\

 }
\label{tb:structural parameters}
\end{table*}

\begin{figure*}
\centering
\includegraphics[height=8.9 cm]{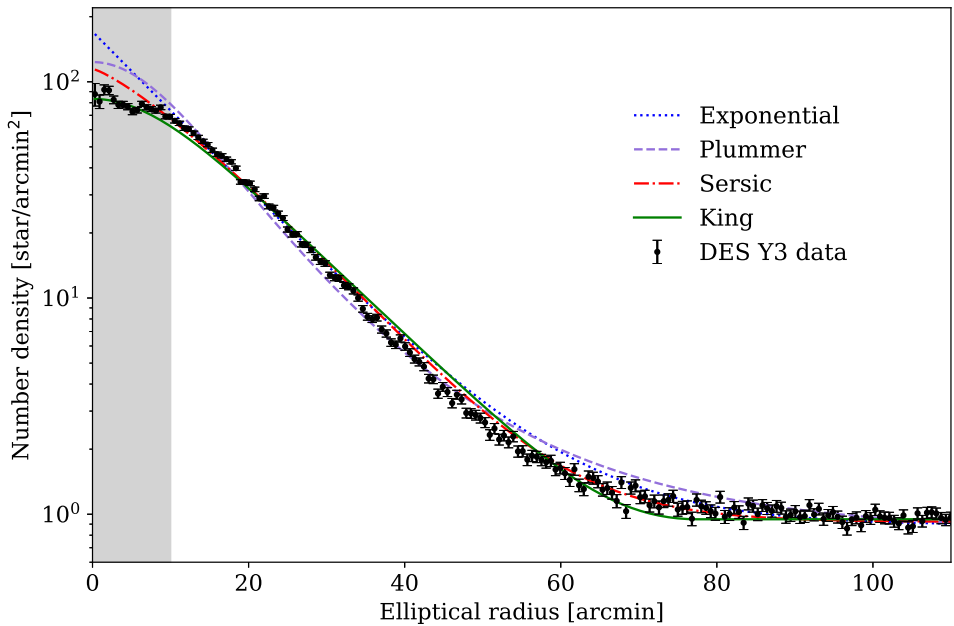}
\caption{The surface density profile for Fornax dSph galaxy with overlaid best-fitting exponential (blue dotted line), Plummer (purple dash line), Sersic (red dash-dotted line) and King models (green solid line). The data is binned in elliptical annuli using the best fit position angle and ellipticity $\epsilon$= 0.31, and the fits were performed to the full unbinned data allowing all parameters to vary freely, as described in Section~\S~\ref{sec:profiles}. We note that our fits are derived excluding stars within 10 arcmin radius (shown with the grey band). Parameter values of the fits are provided in Table~\ref{tb:structural parameters}}
\label{fig:1D_profile}
\end{figure*}

To account for interstellar dust extinction, we followed the procedure described in \cite{DES_2018}. We started with E(B $-$ V) values from the reddening map of \cite{Schlegel_etal1998}. We computed fiducial interstellar extinction coefficients and integrated over the DES standard bandpasses considering a fixed source spectrum that is constant in spectral flux density per unit wavelength. The resulting multiplicative coefficients for the ${\it g}$ and ${\it r}$ band are $R_g$ = 3.185 and $R_r$ = 2.140. Throughout this paper, all magnitudes refer to extinction corrected PSF magnitudes derived by \code{ngmix}.

To assess the crowding issues in Fornex dSph galaxy in the DES photometry, which may affect studies such as galaxy central profile fitting, we implement a pipeline to run \code{DAOphot}~\citep{Stetson_1987} photometry. The \code{DAOphot} approach is designed to better recover stellar sources in crowded fields than \code{SExtractor}, although the drawback is low confidence for star/galaxy separation. The \code{DAOphot} pipeline was implemented for the DES Y3 coadded images~\citep{Morganson_etal18} in \emph{g}, \emph{r} and \emph{i} bands, in a total area of 6.572 square degrees on the sky. After the PSF photometry using \code{DAOphot} tasks was run, the sources were matched using data from 3 bands, with a maximum deviation of 1 arcsec. We also add a zero-point with no color terms. Typical values for the standard deviation of the zero-point are 30 mmag. Similar zeropoints (with a maximum deviation of 0.03 magnitudes) were found comparing to the \code{ngmix} PSF photometry from DES Y3 catalogs. Comparing the sources from the \code{DAOphot} catalog to the DES Y3 sources within magnitude range \textit{g, r, i}=17-21, the magnitude errors are less than 0.03. The final \code{DAOphot} catalog presents a continuous coverage of the sky and successfully recovers the 5 known globular clusters of the Fornax dSph galaxy, for which we find some of them have severe crowding issues in the DES \code{ngmix} photometry. We find mild difference, $\sim 30\%$, in the Fornax central stellar density profile (r $\lsim 10{\mathrm '}$ )between the \code{DAOphot} and the DES \code{ngmix} catalogs, and the comparison is shown in the Appendix. We also find that the \code{DAOphot} catalog is slightly more complete than the ngmix catalog in the inner 10 arcmin of Fornax, though both catalogs still suffer from crowding in this region. For example, we perform artificial star tests for the \code{DAOphot} pipeline on the central tiles of Fornax and assess the completeness level of our catalog. We find that within the central 10${\mathrm '}$ a significant fraction of the input artificial stars are not recovered with the correct magnitude and location due to blending, and that within 10${\mathrm '}$ $<$ 50$\%$ completeness is reached at g$<$ 22 if both criteria of matching spatially with 1${\mathrm "}$ precision and magnitude within 0.5 mag are applied. The incompleteness level decreases for areas further away from the galaxy center, and appears to be symmetric on the East and West sides of Fornax. We therefore mask the inner 10${\mathrm '}$ region when we derive structural parameters. A detailed analysis of the inner region of Fornax is reserved for future work.

\begin{figure*}
\centering
\includegraphics[height=5.2 cm]{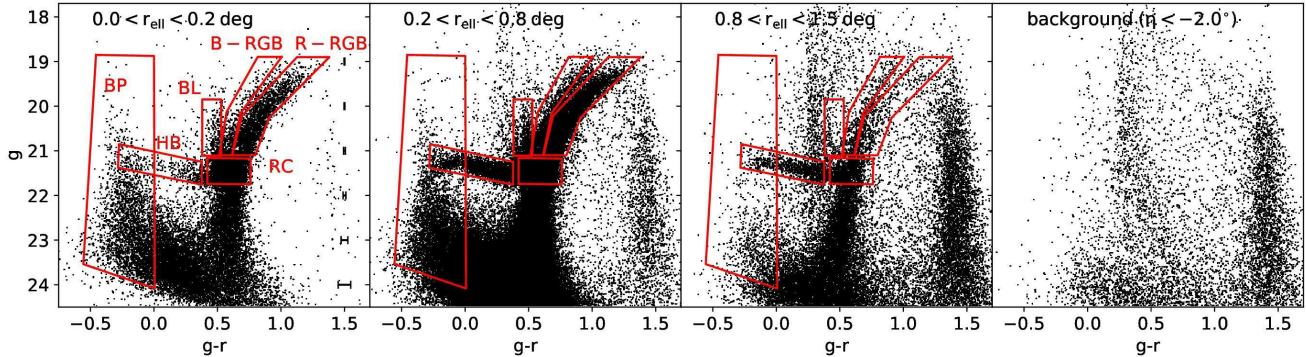}
\vspace{-0.7em}
\caption{Color magnitude diagrams in ${\it g}$ and ${\it r}$ bands for three regions around Fornax and one background region (right most panel). From left to right, we show all stars within an annulus with inner and outer elliptical radius $r_{ell}$ of : $r_{ell} < 0.2^{\circ}$, $0.2< r_{ell} < 0.8^{\circ}$, $0.8< r_{ell} < 1.5^{\circ}$, and the background region in a rectangle with -$2.5^{\circ}$ $<$ $\xi$ $<$ -$2.0^{\circ}$ and -$2.0^{\circ}$ $<$ $\eta$ $<$ $2.0^{\circ}$, where $\xi$ and $\eta$ are coordinates centered on the Fornax centroid. Several different stellar populations are marked by red boxes :
blue red-giant-branch (B-RGB), red red-giant-branch (R-RGB), red clump (RC), horizontal branch (HB), blue plume (BP), and blue loop (BL) stars. The black points with error bars show the average photometry errors at each magnitude bin.}
\label{fig:CMD}
\end{figure*}

\section{STRUCTURAL PARAMETERS}
\label{sec:profiles}

Several previous works have analyzed the radial distribution of stars in Fornax \citep{IH95, Battaglia_etal06, Bate_etal15, del_Pino_etal2015}. Profile models can be fitted to provide important constraints on the size and mass of the galaxy. We re-derive the structural parameters of Fornax dSph using star objects from the uniform, wide coverage DES Y3 data to compare with previous values. We use several different models including an empirical King profile \citep{King_1962}, an exponential profile, a Sersic profile \citep{Sersic_1968}, and a Plummer profile \citep{Plummer_1911}. We fit a surface density profile to the data in a region of radius ${\rm 2.5^{\circ}}$ around the center of the galaxy. Magnitude cuts at ${\it g,r}$ $<$ 23.5 are applied to ensure good quality photometry, and no color cuts are applied. We define the ellipticity $\epsilon$ to be $\epsilon$ = 1 - b/a, where b and a are the semi-minor and semi-major axes of the galaxy respectively. The position angle is taken from North towards East. We use the 2D unbinned maximum likelihood algorithm described in \cite{Martin_etal2008}. The likelihood is evaluated with an affine-invariant ensemble sampler for Markov chain Monte Carlo (MCMC) of the form proposed by \cite{Goodman_etal2010}, which is implemented as the MCMC Hammer \code{emcee} package \citep{Foreman-Mackey_etal2013} for \code{Python}. We allow the following parameters free to change: the center of Fornax (right ascension and declination), the position angle, the ellipticity, the contamination stellar density, and the parameters for the radial density profile. We ran the MCMC with at least an ensemble of 50 walkers, each of which did approximate 1000 steps to ensure chain convergence. We note that the central stellar distribution, as discussed in Appendix, have crowding issues. To reduce the potential effects from crowding on derived structural parameters, we exclude stars within inner 10${\mathrm '}$ radius at the galaxy center while we fit the observed data.  

The structural parameters from fitting various profiles are shown in Table~\ref{tb:structural parameters}, where the uncertainties represent the marginalized errors of 16 and 84 percentile in the fitting procedure. The results from \cite{Battaglia_etal06} and \cite{Bate_etal15} are also shown here for comparison, and are in general agreement with parameters derived here. However, we note a few specific differences when we compare with results from those previous works. For example our fits show systematically noticeable larger values in the effective radius, or the derived 2D half-light radius in all profile types. For example, the resulting Sersic 2D half-light radius $r_h$ is higher by $1.5{\mathrm '}$ (7.5$\%$) than the derived \cite{Battaglia_etal06} value and $2.9{\mathrm '}$ (14.4$\%$) than the \cite{Bate_etal15} result. This is likely due to deeper and better quality of photometry in our data that has resolved more fainter and older main-sequence (MS) population that has more diffused stellar distribution. Our King tidal radius ($r_t$=77.5${\mathrm '}$) is also 7.8${\mathrm '}$ (8.4${\mathrm '}$) larger than the \cite{Bate_etal15} \citep{Battaglia_etal06} values. Otherwise, our position angle, ellipticity, and other structural parameter values are consistent with differences between the literature values.

Figure~\ref{fig:1D_profile} provides the best fit models for each profile compared to the data binned in elliptical annuli with the best-fit position angle and ellipticity. These profiles were obtained using the parameter values from the unbinned fits in Table~\ref{tb:structural parameters}. We also compute the reduced chi-square $\chi^2_{reduced}$ as a goodness-of-fit indicator. It is calculated from the observed surface number density profile derived from elliptical, concentric annuli ranging from 10${\mathrm '}$ to 120${\mathrm '}$ from the galaxy center and the best-fitting profiles, by assuming that star counts follow Poisson distributions with means equal to the star counts from the best-fit model. Similar to \cite{Battaglia_etal06} and \cite{Bate_etal15}, we also found that a Sersic profile provides a much better fit around the tidal radius, whereas the King profile generates slightly better fits at small radii. Overall a Sersic profile provides a best fit at both the tidal radius and in the inner regions, as reflected by smaller $\chi^2_{reduced}$ comparing to other models. We note that from the $\chi^2_{reduced}$ values it is indicated that most of the profile models we consider here are not very good fits to the Fornax light profile. As we will shown later in Section~\S~\ref{sec:MF} and Section~\S~\ref{sec:sub distribution} that near the galaxy center many substructures are present and the inner distribution is slightly off-centered for some stellar populations, and those results in complex profile shape for Fornax that may not be captured by simple models.  
  
\begin{figure*}
\centering
\includegraphics[height=16.0 cm]{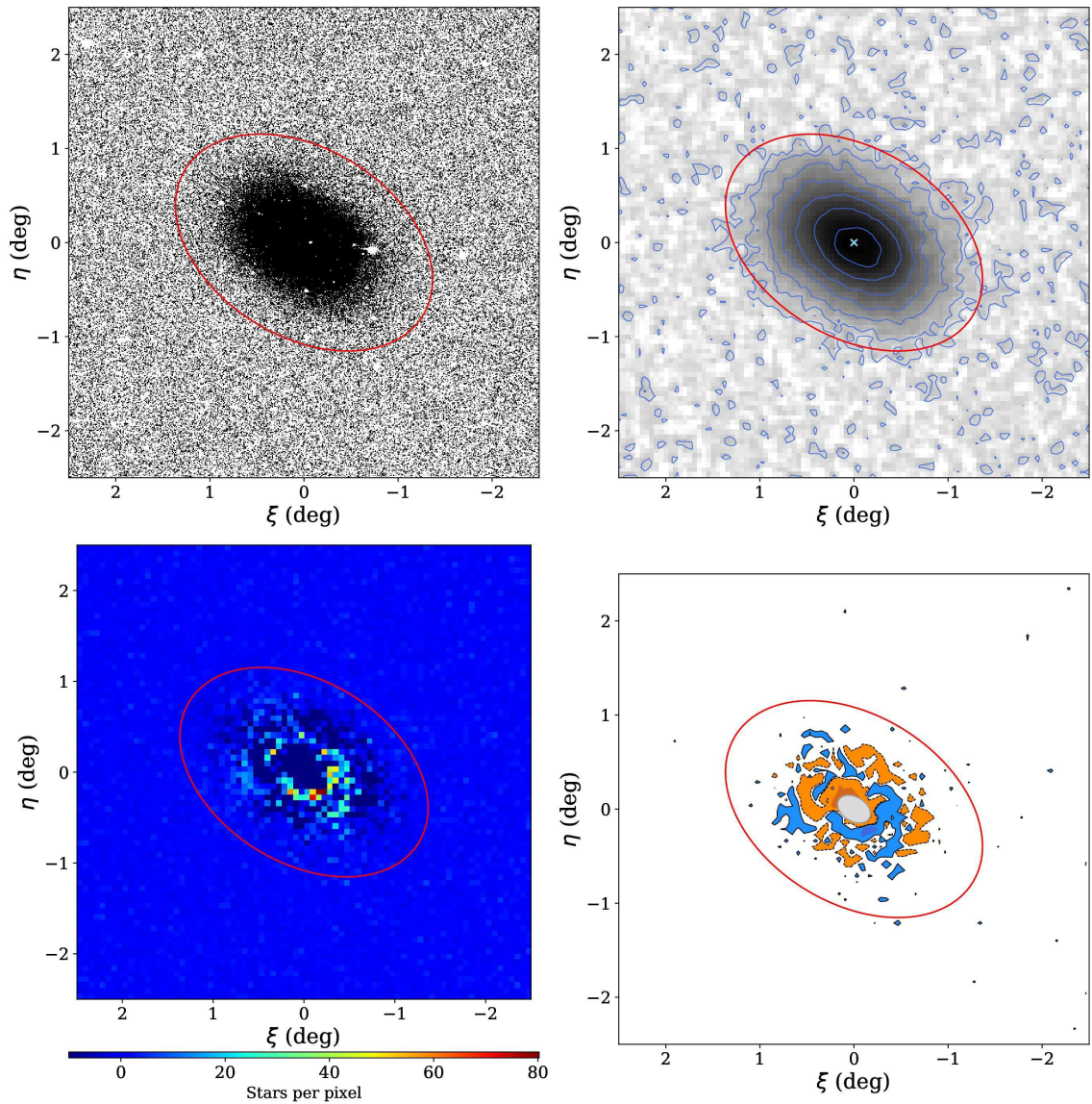}
\caption{\textit{Upper left panel}: spatial distribution of Fornax dSph matched filtered stars. \textit{Upper right panel} :
surface number density map of the matched filtered stars overlaid with iso-density contours. The light blue cross marks the galaxy centroid. \textit{Lower left panel}: Map of residuals between the Fornax surface number density profile from the matched filtered stars and the surface density from the best-fitting Sersic model (see Table~\ref{tb:structural parameters} for the parameter values). Pixel size is ${\rm 2{\mathrm '} \times 2{\mathrm '}}$, smoothed by a Gaussian kernel with 4${\mathrm '}$ dispersion. \textit{Lower right panel} : The residual map with blue (orange) and dark blue (dark orange) contours showing 2$\sigma$ and 3$\sigma$ detections above (below) the mean residuals. The red ellipses in each panel show the nominal King tidal radius derived from this work. The gray ellipse at the center with radius of 10${\mathrm '}$ marks the region that is excluded in the residual analysis.}
\label{fig:MF_spatial}
\end{figure*}

\section{The color-magnitude diagrams of Fornax}
\label{sec:CMD}

In Figure~\ref{fig:CMD} we show ${\it g},{\it g}-{\it r}$ CMDs of several regions of Fornax and one background region. From left to right, we show all stars within an annulus with elliptical radius $r_{ell}$ (with ellipticity = 0.31) of: $r_{ell} < 0.2^{\circ}$, $0.2^{\circ} < r_{ell} < 0.8^{\circ}$ and $0.8^{\circ} < r_{ell} < 1.5^{\circ}$, and a contamination field (right most panel) consisting of data within -$2.5^{\circ}$ $<$ $\xi$ $<$ -$2.0^{\circ}$ and -$2.0^{\circ}$ $<$ $\eta$ $<$ $2.0^{\circ}$, where $\xi$ and $\eta$ are coordinates centered on the Fornax centroid. The CMD of Fornax shows a multitude of features, indicating that it has experienced a prolonged period of star formation. The well sampled stellar features are clearly different from MW contamination features, as traced by the surrounding background region. The bright part of the CMD is dominated by the RGB feature, which shows a clear bifurcation (also see discussions in \cite{Bate_etal15} and \cite{Battaglia_etal06}). Following \cite{Battaglia_etal06} and \cite{Bate_etal15}, we will refer to the two branches of RGBs as the blue RGB (B-RGB) and red RGB (R-RGB). The two features probe stars of different metallicities and ages and show different spatial distribution, pointing to a radial gradient of age within Fornax (see also \cite{Battaglia_etal06} and \cite{de_Boer_etal2012}). Blue loop (BL) stars are also visible around ${\it g-r} \approx$ 0.5, indicating the presence of He-core burning stars with ages ranging from a few million years to 1 Gyr. 
Lower down in the CMD, a strongly populated horizontal branch (HB) and red clump (RC) are visible, both of which change distribution at different radii. Ancient ($>$10 Gyr), metal-poor blue HB stars are strongly present in the $0.8 < r_{ell} < 1.5^{\circ}$ bin, while the RC stars dominate in the innermost bin. This indicates that the radial age gradient is also linked to a metallicity gradient, as also seen from previous spectroscopic measurements at different radii \citep{Battaglia_etal06}. The rich, composite red clump is composed of intermediate age, metal-enriched stars which have been used as a standard candle to determine the distance to Fornax \citep{Bersier_etal00,Rizzi_etal07}.

\begin{figure*}
\centering
\includegraphics[height=12.0 cm]{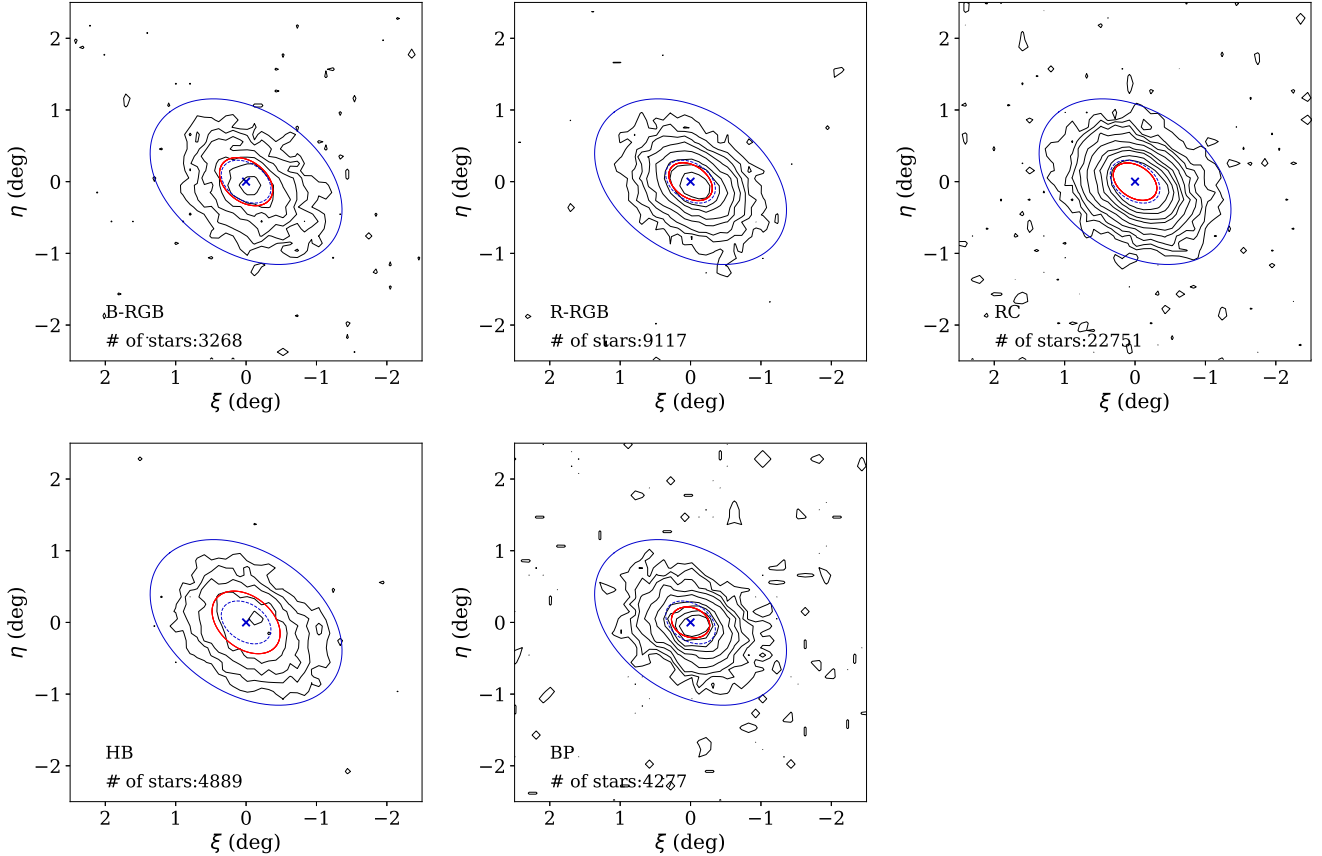}
\vspace{-1.5em}
\caption{The matched filtered stellar density contours for the blue red giant branch (B-RGB, top left panel), red red giant branch (R-RGB, top middle panel), red clump (RC, top right panel), horizontal branch (HB, bottom left panel), and blue plume (BP, bottom middle panel) stars. Contour levels are shown at 10, 25, 50, 100, 200, 400, 600, 1000, 1500 times the root-mean-square values calculated in the contamination region. The blue ellipses show the nominal King tidal radius, and the small blue ellipses with dash lines mark the 2D half-light radius of the overll Fornax population. The red ellipses display the half-light radius from the Plummer profile fits (see Table~\ref{tb:subpopulation_parameter}) of individual subpopulation. The blue crosses mark the galaxy centroid.}
\label{fig:MF_sub_map}
\end{figure*}

The faint part of the CMD is dominated by a composite sub-giant branch and MSTO feature, as a result of multiple overlapping stellar populations covering a large range in age. This is further supplemented by a strongly populated blue plume (BP) population extending to blue color and bright magnitudes. The BP population is composed of genuine young stars, distinct from the population of blue straggler stars which can also be found on the blue-ward side of old MSTOs. The blue stragglers are most visible in the outermost Fornax sample as a sequence of stars extending blueward from the old MSTO, and show a slightly different spatial distribution than the BP stars. Furthermore, the BP stars extend to much brighter magnitudes, only reachable by young ($<$1 Gyr) MS stars. The composition of the faint CMD features also changes at different radii, with younger populations gradually disappearing as the radius increases until only the oldest populations are left over in the outskirts. This once again points to younger stars being preferentially found in the center of Fornax.

The composition of features in the CMD makes it clear that Fornax has formed stars across most of cosmic time without significant interruptions, and just recently ($\approx$ 100 million years ago) stopped star formation \citep{deBoer_etal13}.

\section{Surface density maps}
\label{sec:MF}
The Fornax dwarf galaxy is among the systems most likely to be affected by MW tidal forces, given its relatively cold kinematics for its brightness \citep{Wang_etal17}. If present, these tidal features are likely to have a very low surface density, making them hard to detect against a screen of MW contamination. Therefore, efficient decontamination techniques, such as ``matched-filtering" methods~\citep[e.g.,][]{Rockosi_etal02, Grillmair_etal09}, are needed to enhance their signal over the numerous contaminants.

\begin{table*}
\caption{Fornax Subpopulation Structural Parameters (Plummer model)}
{\renewcommand{\arraystretch}{1.7}
\renewcommand{\tabcolsep}{0.17cm}
\centering
\begin{tabular}{l c c c c c c}
\hline 
\hline
Parameter & B-RGB & R-RGB  & RC & HB & BP\\
\hline
$\Delta \alpha_{2000}$ & ${\rm 0.02^{\circ}}$  & ${\rm 0.01^{\circ}}$ &${\rm 0.005^{\circ}}$ &${\rm 0.014^{\circ}}$  &${\rm -0.003^{\circ}}$  \\
$\Delta  \delta_{2000}$& ${\rm -0.018^{\circ}}$  & ${\rm 0.000^{\circ}}$ &${\rm 0.005^{\circ}}$ & ${\rm 0.015^{\circ}}$& ${\rm -0.003^{\circ}}$\\
Ellipticity& 0.31$\pm$0.01&0.31$\pm$0.01 & 0.29$\pm$0.01  & 0.32$\pm$0.01 & 0.18$\pm$0.01 \\
$\theta$ & $38.9^{\circ}$$\pm$$1.4^{\circ}$ &$42.1^{\circ}$$\pm$$0.8^{\circ}$ & $43.2^{\circ}$$\pm$$0.4^{\circ}$& $37.8^{\circ}$$\pm$$1.1^{\circ}$& $50.6^{\circ}$$\pm$$1.8^{\circ}$ \\
$r_p$ & $22.6{\mathrm '} \pm 0.4{\mathrm '}$ &$18.1{\mathrm '} \pm 0.2{\mathrm '}$ & $18.0{\mathrm '} \pm 0.1{\mathrm '}$& $29.3{\mathrm '} \pm 0.3{\mathrm '}$ & $14.4{\mathrm '} \pm 0.2{\mathrm '}$\\
No. of stars  &3268 &9117 & 22751&4889 & 4277 \\
\hline
\end{tabular}\\

\vspace{0.3em}
 }
\label{tb:subpopulation_parameter}
\end{table*}

\begin{figure}
\centering
\includegraphics[height=7.5 cm]{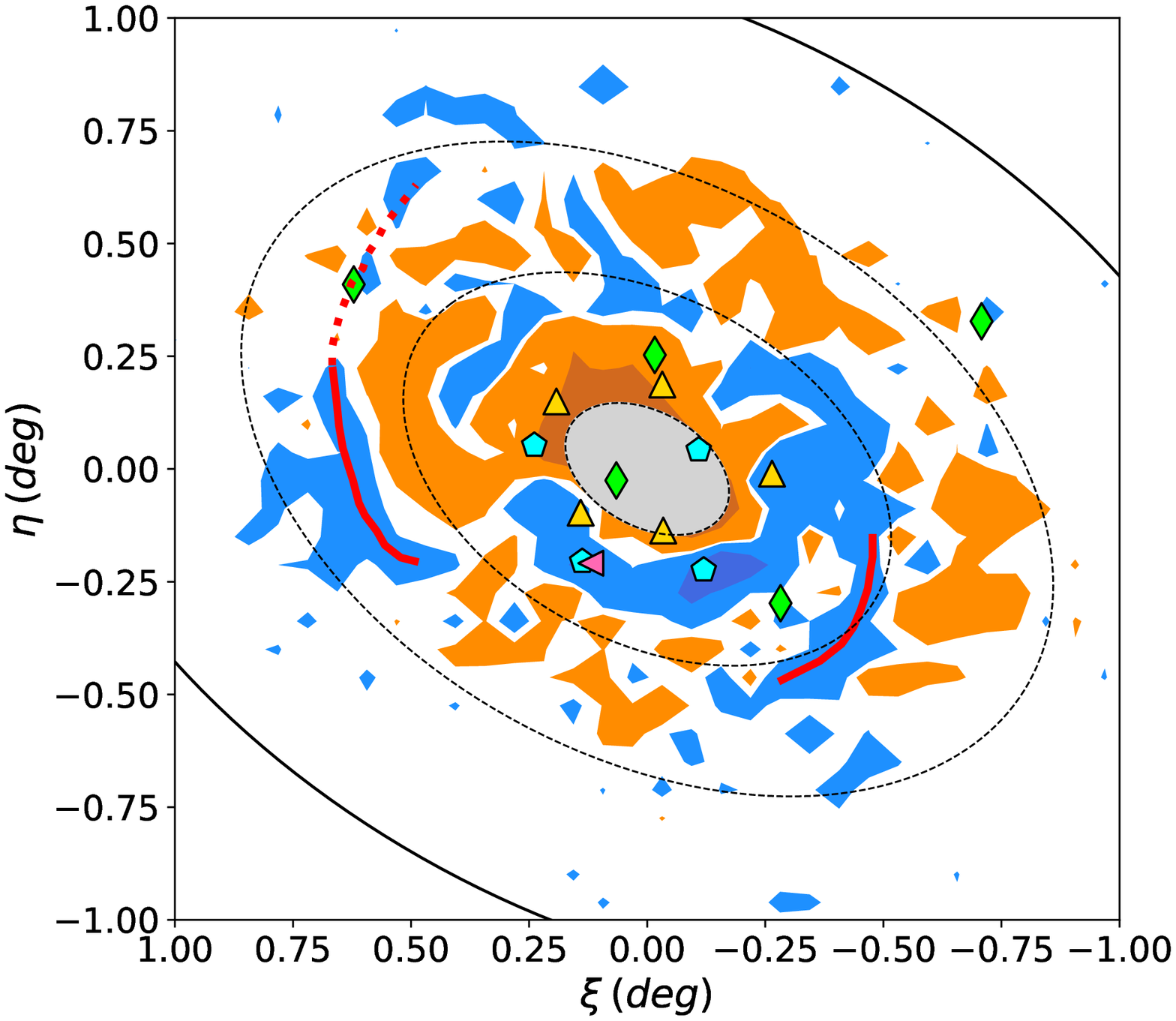}
\vspace{-1.8em}
\caption{The zoomed-in residual map within 4 square degree area of Fornax. The blue (orange) and dark blue (dark orange) contours show the 2$\sigma$ and 3$\sigma$ detection regions above (below) the mean residuals. The black solid ellipses show the tidal radius, and the dashed ellipses are with radii of 0.7, 0.5, and 0.1 times the tidal radius. The light green diamond points show locations of the five known globular clusters, cyan pentagons indicate the position of previously identified overdensities from the literature~\citep{Coleman_etal04,deBoer_etal13,Bate_etal15}, and yellow triangles indicate the position of five significant overdensities from the young star distribution discussed in Section~\S~\ref{sec:overdensity}. The pink triangle marks the location of the shell discussed by \cite{Coleman_etal04}. The red lines on the South-West side and North-East side mark two residual excess regions that resemble possible shell-like features. The red-dash line show the possible extension of the North-East feature that happens to pass through one of the globular cluster. North is up, and East is to the left. The gray ellipse at the center with radius of 10${\mathrm '}$ marks the region that is excluded in the residual analysis.}
\label{fig:MF_residual}
\end{figure}

To investigate the presence of tidal features in Fornax, we construct matched filter maps of different stellar populations. To ensure the filtered data is still Poissonian distributed, we forgo using a classical matched filter technique and instead adopt a Boolean mask procedure. In our case, the Hess diagram of the galaxy dense regions, where the ratio of contamination/source densities is low, is used to build a ``source" filter, defining the shape of the dSph stellar populations in the color-magnitude plane. The source filter region is derived within an ellipse with size of 0.8 times the tidal radius. A ``contamination filter" is obtained from a region far enough to be free of galaxy members, i.e. the background region with -2.5 $< \xi <$ -2.0 and -2.0 $< \eta <$ 2.0 that is shown in the right most panel of Figure~\ref{fig:CMD}.

We describe the source filter, in the form of a probability density function P$_{str}(g,g-r)$ in color-magnitude space. The contamination is described by P$_{bg}(g,g-r)$. Both the source and background filter masks are then normalized, and the fraction P$_{str}(g,g-r)$/P$_{bg}(g,g-r)$ is used to select CMD pixels above a certain threshold. The filter threshold is chosen by finding the value which maximizes the signal-to-noise. We then assign a weight of zero to values of P$_{str}(g,g-r)$/P$_{bg}(g,g-r)$ lower than the threshold and a weight of one for values higher than the threshold.

The decontaminated Fornax stellar spatial distribution derived from the matched-filtering method is displayed in the upper left panel of Figure~\ref{fig:MF_spatial}. In the upper right panel of Figure~\ref{fig:MF_spatial} it is shown in the form of iso-density contours with pixel size ${\rm 3{\mathrm '} \times 3{\mathrm '}}$. There are no significant signs of tidal features or distortion within the 25 square degree area that we have examined. However there are mildly extended low-surface density excesses in the distribution of stars on the East (left hand side in the plot) and the West side of the galaxy, indicating a slightly skewed isophote shape in the outskirts. 

We also derive a two-dimensional map of residuals between the decontaminated surface density map of Fornax stars and the surface density predicted by the best-fitting Sersic model. It is shown in the lower two panels of Figure~\ref{fig:MF_spatial}. Except for some over and under densities in the Fornax inner region that we will discuss later in Section~\S~\ref{sec:overdensity}, it shows reasonable agreement between the decontaminated map of Fornax stars and the models. Based on those maps, we analyze how frequent those residuals appear in the search area and adopt the 1$\sigma$ confidence interval of the distribution as our detection limit; this corresponds to $\sim$1.8 stars/pixel ($\sim$ 32.1 mag/${\rm arcsec^{2}}$). 
We note that the inner 10${\mathrm '}$ region, which is shown as a grey ellipse in the lower right panel of Figure~\ref{fig:MF_spatial}, is excluded in this analysis due to effects from crowding. The lack of significant residuals in the periphery of Fornax dSph galaxy in the residual map therefore indicates no statistically significant tidal debris detection down to a surface brightness limit of $\sim$ 32.1 mag/${\rm arcsec^{2}}$. 

\section{Spatial distribution of different stellar populations }
\label{sec:sub distribution}

The analysis of the spatial distribution of stars in different evolutionary phases is a useful tool to study spatial variations of the stellar population mix as a function of age and/or metallicity. There are multiple examples in the literature of this type of analysis, through which age gradients and star formation history in Fornax dwarf galaxy have been quantified \citep{Battaglia_etal06,Bate_etal15,del_Pino_etal2015}.

In this section, we perform a full matched-filtering method (described in Section~\S~\ref{sec:MF}) and structural analysis of the spatial distribution of Fornax R-RGB, B-RGB, HB, and BP stars. We have chosen not to include an analysis of the RC feature, since the stellar population mix of the RC feature is similar to that of the R-RGB feature. Filters were generated from (${\it g-r}$, ${\it g}$) Hess diagrams, as displayed in Figure~\ref{fig:CMD}. The filters were built by truncating the Hess diagram at appropriate CMD boxes.

\subsection{Matched filtered maps}

The matched filtered maps for four subpopulations are shown in Figure~\ref{fig:MF_sub_map}: the B-RGB, R-RGB, HB, and the BP. Each panel is normalized separately, based on their respective root-mean-square (RMS) values of star counts in the contamination area with -$2.5^{\circ}$ $<$ $\xi$ $<$ -$2.0^{\circ}$ and -$2.0^{\circ}$ $<$ $\eta$ $<$ $2.0^{\circ}$, where $\xi$ and $\eta$ are coordinates centered on the Fornax centroid. The RMS values for each subpopulations are 0.14 star ${\rm pixel^{-1}}$ for B-RGB, 0.14 star ${\rm pixel^{-1}}$ for R-RGB, 0.20 star ${\rm pixel^{-1}}$ for HB, and 0.08 star ${\rm pixel^{-1}}$ for BP. Contour levels are shown at 10, 25, 50, 100, 200, 400, 600, 1000, 1500 times the RMS values calculated in the contamination region. The tidal radius (blue ellipses with solid lines) and the half-light radius (blue ellipses with dash lines) of the overall Fornax distribution derived by this work are plotted in each panel. The red ellipses mark the half-light radius of each subpopulation. The center of the overall Fornax distribution is marked by blue cross point.

\begin{figure}
\includegraphics[width=0.495\textwidth]{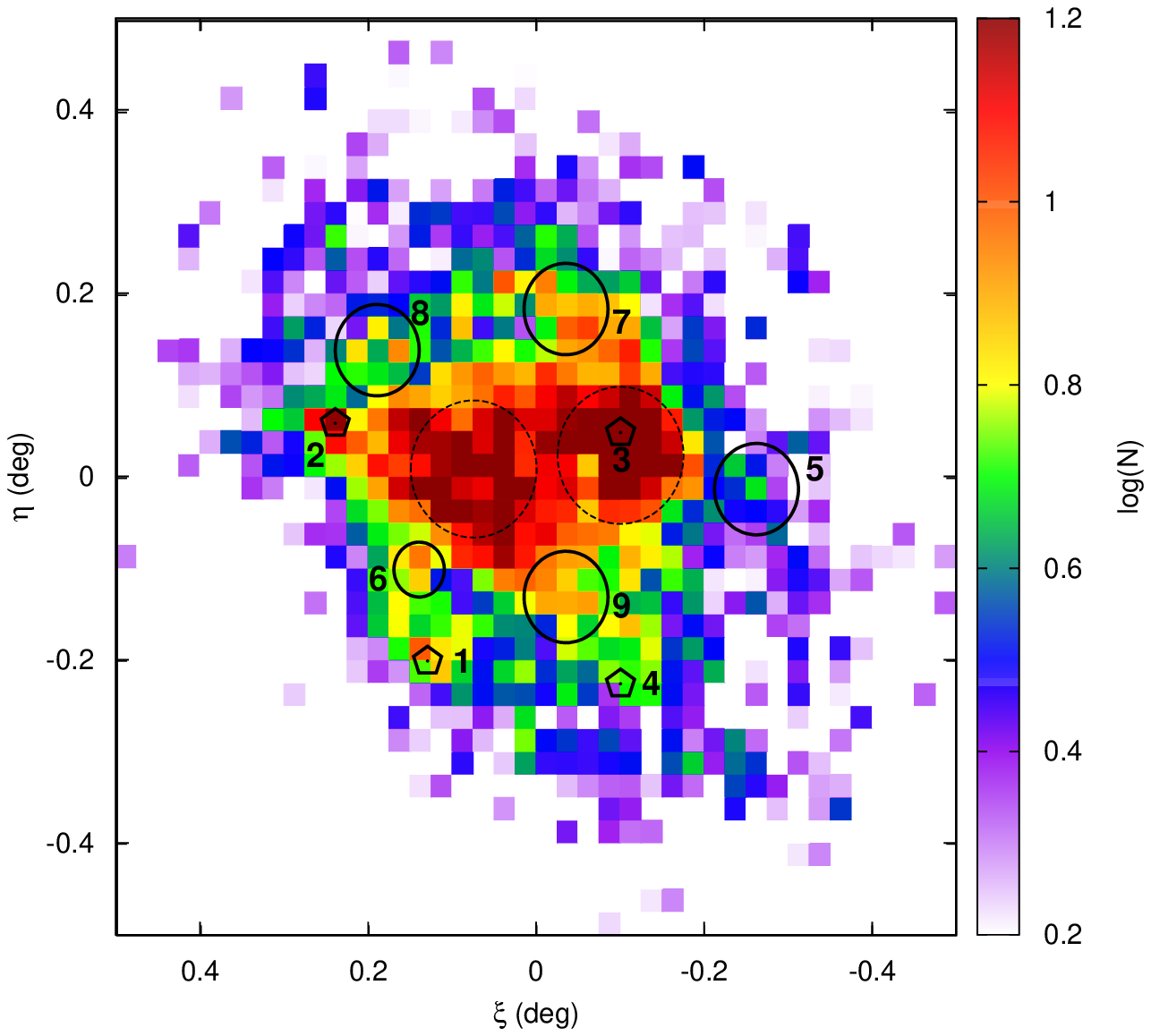}
\vspace{-1.5em}
\caption{The spatial density distribution of Fornax BP stars, selected from the \textit{g} and \textit{r} bands using cuts of \textit{g}$-$\textit{r}$<$0 and r$<$23.5. Black pentagons indicate the position of previously identified overdensities from the literature~\citep{Coleman_etal04,deBoer_etal13,Bate_etal15}. Large dashed circles in the central region indicate the centers of the roughly bimodal distribution making up the core of the young Fornax population. Finally, solid numbered circles indicate the position of five significant overdensities detached from the central young star distribution. North is up, and East is to the left.}
\label{fig:BP}
\end{figure}

\begin{figure}
\includegraphics[width=0.495\textwidth]{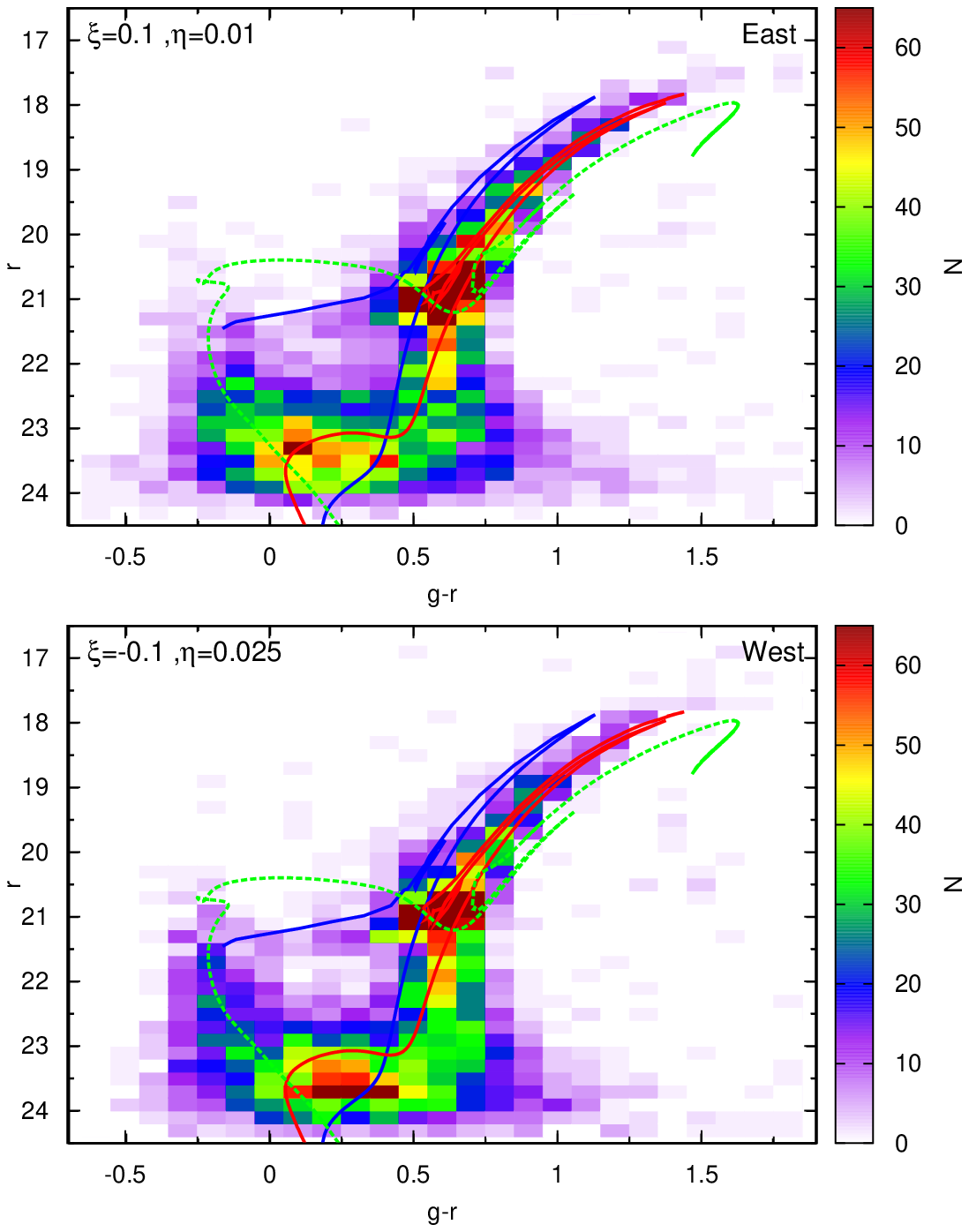}
\vspace{-1.5em}
\caption{The CMDs of the two main over dense regions within the young Fornax populations. These two overdensities show the presence of a wide range of stellar populations, ranging from ancient and metal-poor to intermediate age~($\approx$5 Gyr) stars and young ($<$1 Gyr) metal-rich stars. The blue ([Fe/H]=-2.50 dex, Age=13 Gyr) and red ([Fe/H]=-1.00 dex, Age=4 Gyr) lines show reference isochrones for the main Fornax field population, while the green line indicates a stellar population with solar metallicity at 0.5 Gyr. Comparison between the two panels shows that the Eastern region has younger stars extending all the way up to \textit{r}=19.5 while the young stars in the Western region only extend up to \textit{r}=20.5 and are therefore slightly older.}
\label{fig:centreCMD}
\end{figure}

The B-RGB is more diffuse than the R-RGB, as pointed out in \cite{Bate_etal15}. B-RGB, B-RGB, and RC distributions appear to align with the overall Fornax orientation, but show slightly off-center positions and are slightly more extended toward the South-West side (lower right corners in Figure~\ref{fig:MF_sub_map}). The HB is more diffuse than other subpopulations and shows a mildly extended distribution toward the South-West direction as well. For BP, it is clear that the orientation and the ellipticity of the iso-density contours change to be more aligned with the right ascension direction and become more circularized toward the center. This is also shown in \cite{del_Pino_etal2015}, where it is argued that the young MS stars do not follow the overall Fornax star distribution both in terms of orientation and ellipticity. 

\subsection{Structural parameters}
\label{subsec:subpop profiles}
In order to statistically analyze the compatibility between the spatial distributions of stars in the different evolutionary phases, we obtained their structural parameters by fitting with Plummer models. Since our aim here is not to determine which functional form best represents the surface relative density of the various populations, but to quantify relative differences in their spatial distribution, we fit only a Plummer profile, in order to restrict the number of free parameters. Even in the regime of low number statistics for some of the Fornax subpopulation stars (e.g. B-RGB), the affine-invariant ensemble sampler method provides relatively well-constrained position angle, ellipticity and half-light radii estimates. We note that unlike the overall structural parameter fits in Section~\S~\ref{sec:profiles}, we do not mask the inner 10${\mathrm '}$ for the subpopulation analysis here. Most of the stellar subpopulations we consider here are brighter than g$<$ 22 and therefore subject much less to the incompleteness due to crowding. The faintest subpopulation, BP, actually has the most compact distribution (see the discussion below). This is opposite of what would expect from a completeness artifact. Still this may impose greater uncertainties on the BP structural parameters than the values quoted here, and readers should proceed with caution with those results. 

\begin{figure*}
\includegraphics[width=0.95\textwidth]{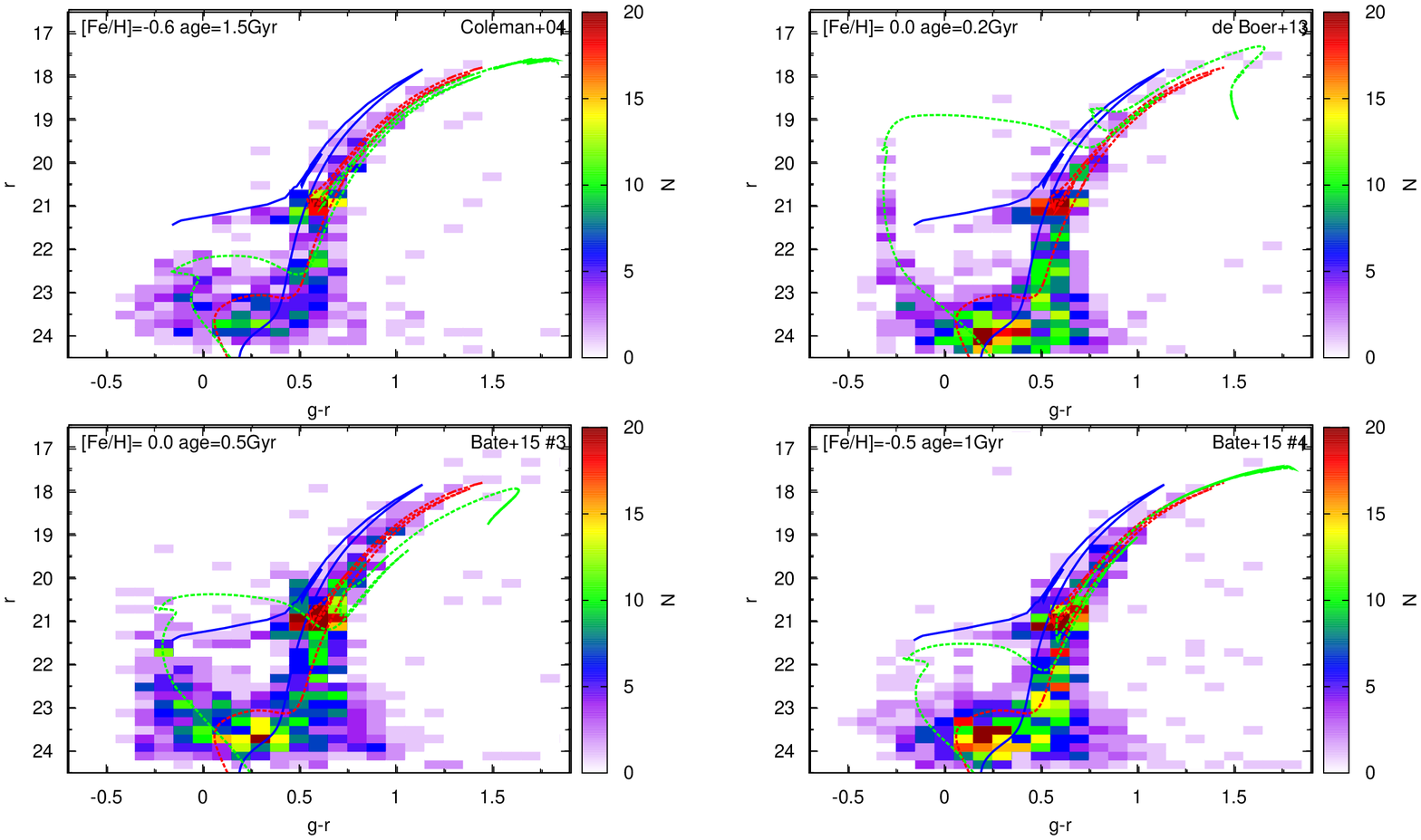}
\vspace{-1.5em}
\caption{The CMDs of previously identified overdensities in Fornax from the literature (\#1 is from \cite{Coleman_etal04}, \#2 from \cite{deBoer_etal13}, and \#3, 4 are from \cite{Bate_etal15}). Overlaid on the CMDs are reference isochrones for the Fornax field population~(blue: [Fe/H]=-2.50 dex, Age=13 Gyr, red: [Fe/H]=-1.00 dex, Age=4 Gyr) as well the locus of the main overdense populations~(in green with metallicity and age marked in each panel).}
\label{fig:knownOD}
\end{figure*}

In Table~\ref{tb:subpopulation_parameter} we summarize the Plummer model fitting results of these various subpopulations. The results listed in Table~\ref{tb:subpopulation_parameter} reflect what are shown in the matched filtered contour maps in Figure~\ref{fig:MF_sub_map} and provide quantitative descriptions. The red ellipses in each panels in Figure~\ref{fig:MF_sub_map} also show the orientation, ellipticity, and the half-light radius from the Plummer model fits. The centroids of all four subpopulations show very mild deviations from the Fornax center. It is clear in Table~\ref{tb:subpopulation_parameter} the HB has a larger Plummer radius $r_p$ ($r_p$=29.3${\mathrm '}$) than other subpopulations and the overall population ($r_p$=19.4${\mathrm '}$). The BP population is the most compact (with the smallest $r_p$ as 14.4${\mathrm '}$) and has a much rounder shape (ellipticity=0.18) than other populations. The R-RGB has profile parameters most similar to the overall Fornax population (see Table~\ref{tb:structural parameters}). For B-RGB and RC, they are also similar to the overall Fornax population except for a slightly different orientation or ellipticity.

\section{Inner overdensities}
\label{sec:overdensity}

In Figure~\ref{fig:MF_residual} we show a zoomed-in version of the residual map (the same as the lower right panel of Figure~\ref{fig:MF_spatial}) within the 4 square degree area around the Fornax center, and we mark several previously known and newly discovered features by this work. The blue (orange) and dark blue (dark orange) contours respectively show the 2$\sigma$ and 3$\sigma$ detections above (below) the mean residuals. The light green diamond points show locations of the known globular clusters, cyan pentagons indicate the position of previously identified overdensities from the literature~\citep{Coleman_etal04,deBoer_etal13,Bate_etal15}, and yellow triangles indicate the position of five new overdensities in the young star distribution discussed in Section~\S~\ref{sec:YS overdensity}.

We also mark two regions of residual excesses with red lines on the South-West side and North-East side that resemble shell-like features wrapping around the galaxy. These low surface brightness features are $\sim 30-40^{\prime}$ away from the galaxy center. The nature and origin of these features is unclear, but there are hints in the literature that suggest Fornax has experienced one or more merger events in the past. The features presented here could be debris of those tidally disrupted stellar systems. For example, this scenario is supported by chemo-kinematic analyses of Fornax, which reveal that Fornax may have complex dynamical patterns including multi rotation components with different metallicity distributions or directions of angular momentum \citep{Amorisco_etal2012,del_Pino_etal2017}. This is ascribed to a merger between two or more stellar systems that each carry a particular angular momentum when they merged with Fornax. Furthermore, several stellar substructures have been linked to possible merger origins \citep{Coleman_etal04,Yozin_etal2012,del_Pino_etal2015}. If some of those globular clusters in Fornax are brought in by merger events in the past, this might explain Fornax's unusually high number of globular clusters.

\subsection{Young stellar overdensities}
\label{sec:YS overdensity}
The inner regions of Fornax (i.e. r$_{ell}<0.5^{\circ}$) are dominated by stars of intermediate age, and show a large fraction of stars younger than $\approx$2 Gyr on the young MS. These BP stars are readily visible in Figure~\ref{fig:CMD} for blue colors with ${\it g-r}<$0 extending up to ${\it g} \approx$20. Previous studies of the young stars have discovered several overdensities, some of which are composed of stars younger than what are found in the Fornax field~\citep{Coleman_etal04,deBoer_etal13,Bate_etal15}.

\begin{table}
\caption{The model adopted for the age and metallicity of the Fornax young stellar overdensities}
{\renewcommand{\arraystretch}{1.3}
\renewcommand{\tabcolsep}{0.16cm}
\centering
\begin{tabular}{c c c c c c c}
\hline 
\hline
Overdensity & Age (Gyr) & [Fe/H] & Reference\\
\hline
1 & 1.5 & -0.6 & \citet{Coleman_etal04}\\
2 & 0.2 & 0.0 & \citet{deBoer_etal13}\\
3 & 0.5 & 0.0 & \citet{Bate_etal15}\\
4 & 1.0 & -0.5 & \citet{Bate_etal15}\\
\hline
5 & 1.5 & -0.6 & This Work\\
6 & 1.5 & -0.6 & This Work\\
7 & 0.5 & 0.0 & This Work\\
8 & 0.2 & 0.0 & This Work\\
9 & 1.0 & -0.5 & This Work\\
\hline
\end{tabular}\\
\vspace{0.3em}
 }
\label{tb:ODs}
\end{table}

\begin{figure*}
\includegraphics[width=0.95\textwidth]{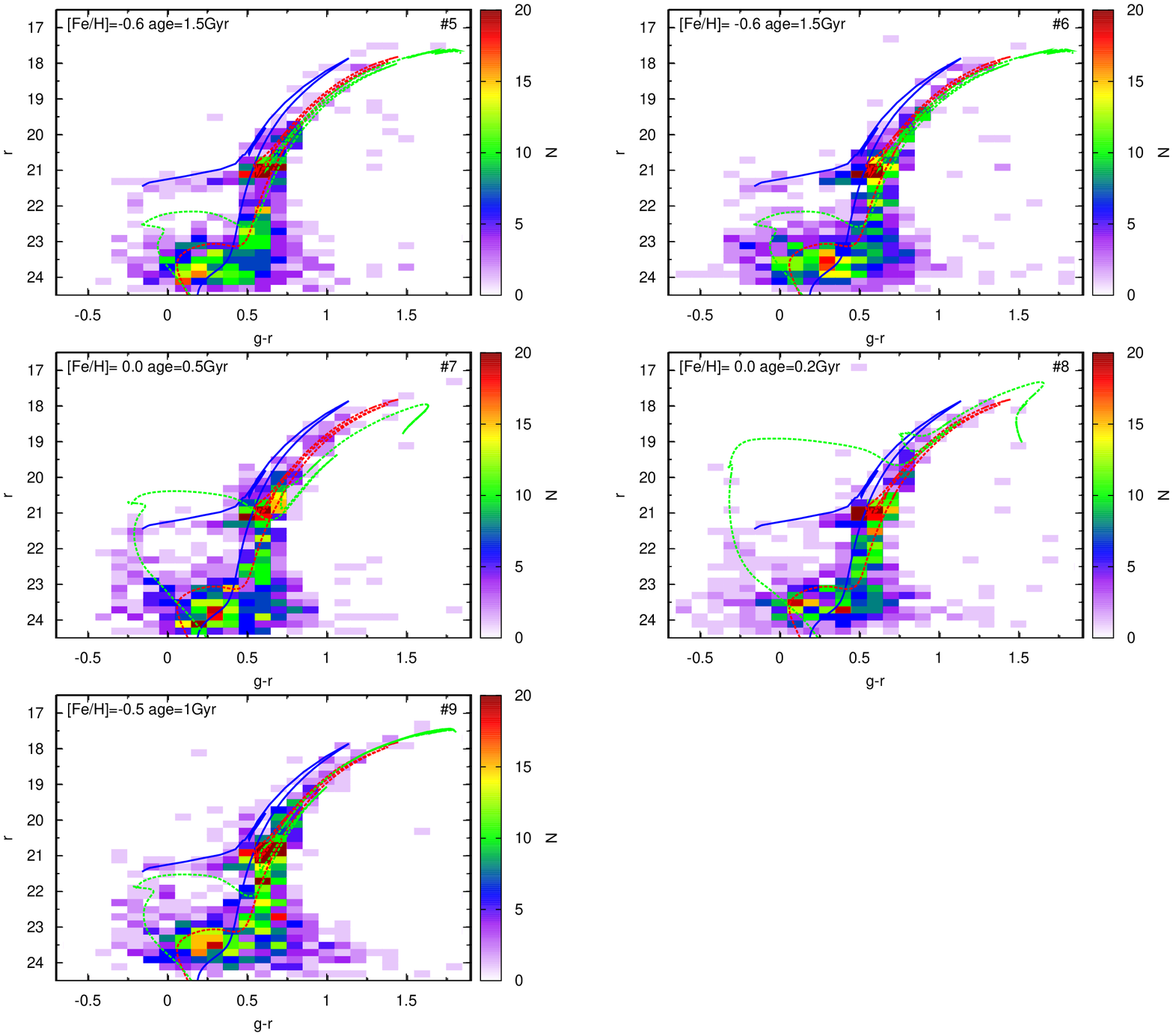}
\vspace{-1.5em}
\caption{The CMDs of the five significant overdensities in Fornax young stars shown in Figure~\ref{fig:BP}. Once again, reference isochrones for the Fornax field population~(blue: [Fe/H]=-2.50 dex, Age=13 Gyr, red: [Fe/H]=-1.00 dex, Age=4 Gyr) as well the locus of the main over dense populations~(in green with metallicity and age marked in each panel) are overlaid.}
\label{fig:newOD}
\end{figure*}

To study the intermediate age and young population in DES Y3 data, we select stars on the BP region with a color cut shown in Figure~\ref{fig:CMD}. We exclude stars fainter than ${\it g}$, ${\it r}$=23.5 to avoid stars on the old MS which may scatter into our selection box. In Figure~\ref{fig:BP}, we show the spatial density of BP stars, with colors representing the density in each bin. The first thing to notice in Figure~\ref{fig:BP} is that the young stars are oriented quite differently than the old stellar populations on the RGB, which is also shown in Figure~\ref{fig:MF_sub_map} and the position angle from the Plummer model fits (see Table~\ref{tb:subpopulation_parameter}). The central distribution within $0.2^{\circ}$ is aligned very close to the right ascension direction, as pointed out in Section~\S~\ref{sec:sub distribution}. However, there is a diffuse and extended component that is more aligned with overall orientation and ellipticity of Fornax, and therefore the derived position angle for BP stars in Table~\ref{tb:subpopulation_parameter} only differs by $7.8^{\circ}$ comparing to the overall Fornax position angle.The origin of this orientation mismatch is unclear, but may be linked to the episodic formation history of Fornax and the possibility of gas accretion fueling the creation of the young stars~\citep[e.g.,][]{Yuan_etal16}. The distribution of young stars can be roughly characterized as two large overdensities on either side of the North-South axis, with the Eastern one being slightly more extended.

To further investigate the central populations, we extract the CMD of the central features within the large dashed circles shown in Figure~\ref{fig:BP}. These two circles are positioned on the peaks of the two main density distributions, after doing a simple gaussian fit. For comparison, we show the expected distribution of young, intermediate and old stars by overlaying stellar isochrone from the MIST isochrone library~\citep{Dotter_etal16, Choi_etal16}. The CMDs show the presence of a wide range of stellar populations, including ancient metal-poor stars on the blue RGB and the dominant intermediate age ($\approx$5 Gyr), metal-rich population that results in a red RGB and dense red clump. Furthermore, as expected, the CMDs show a clear presence of young stars on the BP population. The density and morphology of the young BPs is different between the two regions, with stars in the Eastern region~(upper panel) extending to brighter magnitudes. Comparison to stellar isochrones, which is shown in Figure~\ref{fig:centreCMD}, indicates that the Western region is well matched to a population with solar metallicity at an age of 0.5 Gyr, while the Eastern region has brighter stars corresponding to a younger MS age. 

Figure~\ref{fig:BP} shows several other overdense features of young stars surrounding the central features. The positions of known over densities are labeled by pentagons, and clearly correlate with regions with a density higher than the local background. The known overdensities are described in more detail in~\citet{Coleman_etal04} (\#1, bottom left pentagon), \citet{deBoer_etal13} (\#2, top left pentagon), and~\citet{Bate_etal15} (\#3 and \#4, top right and bottom right pentagons respectively). We extract stars of the known features using regions with a radius of 2$^{\prime}$ and show the resulting CMDs in Figure~\ref{fig:knownOD}. Comparing to the VST/ATLAS data from~\citet{Bate_etal15}, the DES Y3 photometry is deeper and allows us to characterize the stellar populations of these features with greater accuracy. Therefore, we determine the best fit to the over dense population using a modified version of the \code{Talos} code~\citep{de_Boer_etal2012} which fits a single isochrone (with a fixed AMR) to a cut-out of the CMD centered on the blue part of the CMD. These fits are made to generate an indication of the age of the overdensities and no attempt is made here to fit the detailed shape of the density distribution. The results of the fits are summarized in Table~\ref{tb:ODs}. Metallicities are assumed to follow the Fornax young age-metallicity relation as determined in~\citet{deBoer_etal13} for the field population. Given their brightness, no spectroscopic metallicities are known for the young stars of Fornax. We find that the overdensity discovered by \citet{Coleman_etal04} (\#1) is the oldest with an age of 1.5 Gyr in good agreement with the literature. The two features discovered by~\citet{Bate_etal15} (\#3 and \# 4) have ages of 0.5 and 1 Gyr respectively, while the overdensity found previously by~\citet{deBoer_etal13} (\#2) has a very young age of 0.2 Gyr. It is interesting that the youngest overdensities are found on the Eastern side of Fornax, for which the field population is also younger than the Western side. This might be an indication that this side of the Fornax stellar distribution might be younger as a whole, possibly as the result of an external event.

The distribution of young stars in Figure~\ref{fig:BP} shows several other high density regions away from the main central core. These might simply be extensions of the main Fornax field distribution, but could also be separate structures with a distinct population. To that end, we select five high density regions highlighted in Figure~\ref{fig:BP} and study their CMDs. Those regions are placed on local overdense peaks with N $>$10 per pixel that are sufficiently far from previously known high overdensities. The location is defined using a simple Gaussian fit to the region surrounding the hot pixel. We include region \#5 in our selection despite not showing a greatly increased density, due to its location on the opposing side of the very young known overdensity. 

Due to the irregular morphology of the Fornax young star distribution, it is not possible to model and subtract the young field distribution and determine the size of the overdensities. Instead, we select all stars around high density features within a radius of 3$^{\prime}$ as shown in Figure~\ref{fig:BP}. For overdensity \#6 we adopt a smaller radius of 2$^{\prime}$ to avoid overlapping with the main Fornax field population. Figure~\ref{fig:newOD} shows the CMDs of these five regions, along with reference isochrones for the field population. The over dense young populations all show blue colors, well below those of the MW foreground (which mainly occupy ${\it g-r}>$0.2). Therefore, foreground contamination is negligible apart from the possibility of MW white dwarfs, which are not numerous in the small spatial areas probed here and do not result in the same CMD morphology.

The CMD of overdensity \#5 makes it clear that the overdensity is mostly a result of blue horizontal branch stars within our BP selection box, rather than genuine young stars. Some young stars with ages of $\approx$1.5 Gyr are present, but not in great number. Therefore, we conclude this feature is not a young over density, highlighting even more the asymmetric nature of the young star distribution. The CMDs of the other new high density regions all show an overabundance of young stars with morphologies well traced by the isochrones. Comparison between the new and old overdensities shows that the populations of \#6 and the shell of \citet{Coleman_etal04} are consistent, indicating a possible link between them. A possible link between region \#4 in \citet{Bate_etal15} and our region \#9 is also plausible, given that its populations match quite well, and are different from those found in the Fornax center. The very young stars in \#8 are also similar to those found in the nearby over density of \citet{deBoer_etal13}. Finally, the strong overdensity \#7 has populations comparable to those found in the Eastern central region. It is therefore possible that this feature is simply an extension of the young Fornax field population to the North. 

\section{Conclusions and Discussions}
\label{conclusions}

In this work we utilize the DES Y3 data that has homogeneous wide-field coverage of the Fornax dSph galaxy down to a depth of $g, r \sim 23.5$ to study several properties of the galaxy. 

We have fitted the Fornax structural parameters with several different model including Plummer, Sersic, Exponential, and King profiles, and found that Sersic model gives the best fit to the data while other models seem to fail to capture the complex light profile shape of Fornax. We also found that our data indicates Fornax is more extended than what was shown in previous studies. For example our results show slightly larger half-light radius than the literature values, which were derived either from shallower data or incomplete coverage. Our derived King tidal radius $r_t$, which has the value of 77.5${\mathrm '}$, is also larger than previous results ($r_t$=69.7${\mathrm '}$ from \cite{Bate_etal15} and $r_t$=69.1${\mathrm '}$ from \cite{Battaglia_etal06}). Nevertheless, in general the rest of the properties such as ellipticity, position angle, profile characteristic radii agree with previous results.

To examining the possible low surface density tidal feature around Fornax, we also applied the matched filtering method to enhance the signal over contamination. We examined 25 square degree area around Fornax, and found no significant tidal features or distortion down to $\sim$ 32.1 mag/${\rm arcsec^2}$ with 1$\sigma$ confidence interval by analyzing the frequency distribution of the smooth model subtraction residuals. 

We further examine the spatial distribution of several different stellar populations including R-RGB, B-RGB, HB, and BP stars by checking their matched filtered maps and fitting with Plummer profile model. We find that the HB stars are more spatially extended than the RGB stars, in accordance with previous findings. Our matched filter maps of R-RGB, B-RGB, and BP stars show mild departure from symmetric distribution and have distorted or extended features toward the South-West side. The orientation of BP stars in the center region deviates significantly from the overall Fornax orientation and gradually changes further out from the center. 

When we subtract a smooth Sersic model from the observed stellar distribution and check the distribution of residuals, we find two low surface density features. These shell-like features are located $\sim 30^{\mathrm '}-40^{\mathrm '}$ away from the galaxy center and wrap around Fornax at the North-East side and South-West side. Those features are likely debris due to past merger events.

We have studied the young populations in Fornax by focusing on the distribution and morphology of blue stars (${\it g}-{\it r}<$0) on the young MS. Stars that occupy this region of the CMD have ages typically below 2 Gyr, and can be used to study the spatial distribution of Fornax in a regime where previous studies have seen hints of interactions or accretion~\citep{Coleman_etal04, Coleman_etal05b}. The depth of DES Y3 data allows us to study the stellar population content in young stars with greater accuracy than was previously possible using other wide-field surveys such as VST/ATLAS.

The spatial density of young stars (see Figure~\ref{fig:BP}) shows a very different morphology than the distribution of old stars in Fornax. While the dominant 4$-$7 Gyr old stars have a position angle of nearly 45 degrees, the inner young stars are in bimodal distribution that is more aligned along the right ascension direction. This indicates that the change in spatial distribution occurred in a time frame of at most several Gyr. This is a promising opportunity for deciphering the formation history of Fornax, especially given that the {\it Gaia} satellite~\citep{GAIAmain} will produce accurate proper motions for the brightest stars in this sample.

The young star distribution in Figure~\ref{fig:BP} shows several high density regions, some of which are new and some were previously discovered~\citep{Coleman_etal04, deBoer_etal13, Bate_etal15}. We use the deep DES data to assign an age to the over dense population based on their CMD morphology (see Figure~\ref{fig:knownOD}). We find a range of ages between 0.2 Gyr and 1.5 Gyr, in good agreement with literature values. Given the discrete sets of populations we find, the star formation happened in a stochastic way with inhomogeneous mixing across the galaxy. 

It is striking that the youngest over densities are all found on the Eastern side of Fornax, where the field population itself is slightly younger than in the West (see Figure~\ref{fig:centreCMD}). Furthermore, we find tantalizing evidence that the previously discovered overdensities are linked to other over dense features further in towards the bulk of the Fornax young field population. This gives the impression that these features are in fact shells or clumps of stars stripped out from the central young distribution. It is unclear what implications this has for the recent evolution of Fornax, especially in the minor merger scenario~\citep{del_Pino_etal2015}. The morphology and ages of the features imply that Fornax may have undergone an encounter of some sort during the last 2 Gyr which radically changed the spatial distribution of newly formed stars but left the distribution of already formed stars intact. An accurate kinematic study of these relatively faint stars (down to ${\it r} \approx$22.5) would be invaluable in this context.

\acknowledgements{
\textit{Acknowledgements} :  We would like to thank Vasily Belokurov for helpful discussions. MYW acknowledges support of the McWilliams Postdoctoral Fellowship. T.d.B. acknowledges support from the European Research Council (ERC StG-335936). This paper has gone through internal review by the DES collaboration.

This research has made use of NASA's Astrophysics Data System Bibliographic Services.

Funding for the DES Projects has been provided by the U.S. Department of Energy, the U.S. National Science Foundation, the Ministry of Science and Education of Spain, the Science and Technology Facilities Council of the United Kingdom, the Higher Education Funding Council for England, the National Center for Supercomputing Applications at the University of Illinois at Urbana-Champaign, the Kavli Institute of Cosmological Physics at the University of Chicago, the Center for Cosmology and Astro-Particle Physics at the Ohio State University, the Mitchell Institute for Fundamental Physics and Astronomy at Texas A\&M University, Financiadora de Estudos e Projetos, Funda{\c c}{\~a}o Carlos Chagas Filho de Amparo {\`a} Pesquisa do Estado do Rio de Janeiro, Conselho Nacional de Desenvolvimento Cient{\'i}fico e Tecnol{\'o}gico and the Minist{\'e}rio da Ci{\^e}ncia, Tecnologia e Inova{\c c}{\~a}o, the Deutsche Forschungsgemeinschaft and the Collaborating Institutions in the Dark Energy Survey. 

The Collaborating Institutions are Argonne National Laboratory, the University of California at Santa Cruz, the University of Cambridge, Centro de Investigaciones Energ{\'e}ticas, Medioambientales y Tecnol{\'o}gicas-Madrid, the University of Chicago, University College London, the DES-Brazil Consortium, the University of Edinburgh, the Eidgen{\"o}ssische Technische Hochschule (ETH) Z{\"u}rich, Fermi National Accelerator Laboratory, the University of Illinois at Urbana-Champaign, the Institut de Ci{\`e}ncies de l'Espai (IEEC/CSIC), the Institut de F{\'i}sica d'Altes Energies, Lawrence Berkeley National Laboratory, the Ludwig-Maximilians Universit{\"a}t M{\"u}nchen and the associated Excellence Cluster Universe, the University of Michigan, the National Optical Astronomy Observatory, the University of Nottingham, The Ohio State University, the University of Pennsylvania, the University of Portsmouth, SLAC National Accelerator Laboratory, Stanford University, the University of Sussex, Texas A\&M University, and the OzDES Membership Consortium.

Based in part on observations at Cerro Tololo Inter-American Observatory, National Optical Astronomy Observatory, which is operated by the Association of Universities for Research in Astronomy (AURA) under a cooperative agreement with the National Science Foundation.

\begin{figure*}
\centering
\includegraphics[height=6.3 cm]{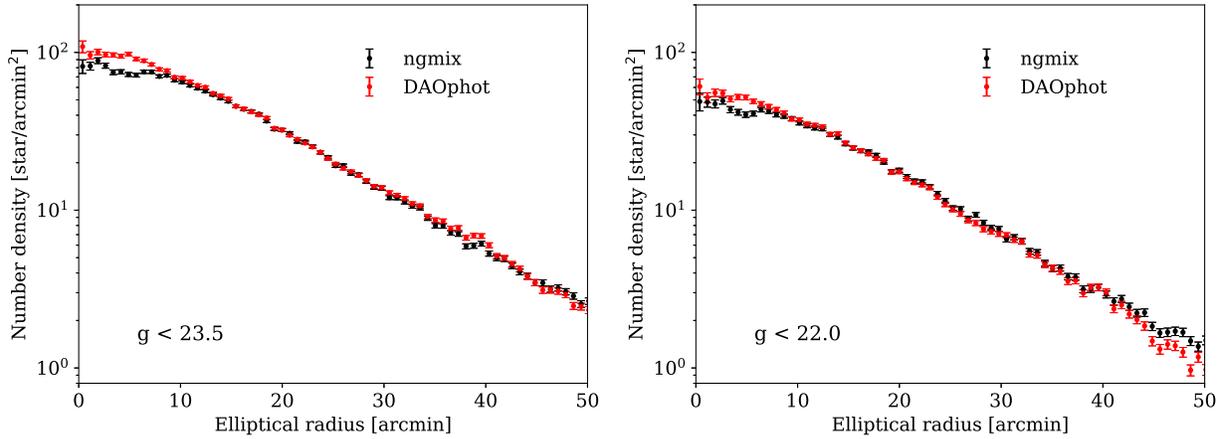}
\caption{$\it{Left}$ $\it{panel}$:the surface density profile for Fornax dSph galaxy with stars of $\it{g<}$ 23.5 overlaid with best-fitting Sersic models from DES \code{ngmix} photometry data (model:black dash-dotted line; data:black points) and \code{DAOphot} data (model:red dash-dotted line; data:red points). The data is binned in elliptical annuli using the best fit position angle and ellipticity, and the fitting procedure are described in Section~\S~\ref{sec:profiles}. $\it{Right}$ $\it{panel}$:the surface density profile from DES \code{ngmix} and \code{DAOphot} data with stars of $\it{g<}$ 22.0.}
\label{fig:daophot_profile}
\end{figure*}

The DES data management system is supported by the National Science Foundation under Grant Numbers AST-1138766 and AST-1536171.
The DES participants from Spanish institutions are partially supported by MINECO under grants AYA2015-71825, ESP2015-66861, FPA2015-68048, SEV-2016-0588, SEV-2016-0597, and MDM-2015-0509, 
some of which include ERDF funds from the European Union. IFAE is partially funded by the CERCA program of the Generalitat de Catalunya.
Research leading to these results has received funding from the European Research
Council under the European Union's Seventh Framework Program (FP7/2007-2013) including ERC grant agreements 240672, 291329, and 306478.
We  acknowledge support from the Australian Research Council Centre of Excellence for All-sky Astrophysics (CAASTRO), through project number CE110001020, and the Brazilian Instituto Nacional de Ci\^encia
e Tecnologia (INCT) e-Universe (CNPq grant 465376/2014-2).

This manuscript has been authored by Fermi Research Alliance, LLC under Contract No. DE-AC02-07CH11359 with the U.S. Department of Energy, Office of Science, Office of High Energy Physics. The United States Government retains and the publisher, by accepting the article for publication, acknowledges that the United States Government retains a non-exclusive, paid-up, irrevocable, world-wide license to publish or reproduce the published form of this manuscript, or allow others to do so, for United States Government purposes.

\textit{Facility:} Blanco (DECam)

\textit{Software:} \code{SExtractor} \citep{Bertin_etal02}, \code{DAOPhot} \citep{Stetson_1987}, \code{astropy} \citep{Astropy}, \code{matplotlib} \citep{Matplotlib}, \code{numpy} \citep{NumPy}, \code{scipy}, \code{ngmix} \citep{Sheldon_14}, \code{emcee}\citep{Foreman-Mackey_etal2013}
}

\appendix
\section{Surface density profile from DAOphot data}
\label{appendix}
To evaluate the impact of crownding on the Fornax DES Y3 data and how much the derived structural parameters could potentially be affected, we generate \code{DAOphot} (see Section~\S~\ref{observations}) catalogs and compare the resulting surface density profiles between these two data-sets. The binned 1D density profiles from DES Y3 data (black points) and the \code{DAOphot} data (red points) are displayed in Figure~\ref{fig:daophot_profile}. In general the DES Y3 profile agrees well with the \code{DAOphot} profile. However, it is noticeable that at the center region (r $<$ 10${\mathrm '}$) the profile amplitude from the DES Y3 \code{ngmix} catalog is slightly lower than the \code{DAOphot} result. The density discrepancy at the center is more prominent for faint magnitude limits. This is shown in Figure~\ref{fig:daophot_profile} when we compare profiles in the left panel (stars with g $<$ 23.5) to the ones in the right panel (stars with g $<$ 22.0). This indicates that the DES Y3 data does present crowding issues at the center of Fornax. Also as mentioned in Section~\S~\ref{observations}, we perform artificial start tests and find that the \code{DAOphot} catalog is $<$50$\%$ complete within the inner 10${\mathrm '}$ for stars with g $<$ 22. In order to reduce the impacts of crowding on the derived structural parameters, we therefore mask the central 10${\mathrm '}$ region when we perform the fits.  

\vspace{1.5em}

\bibliographystyle{apj}
\bibliography{main}{}

\begin{thebibliography}{}
\expandafter\ifx\csname natexlab\endcsname\relax\def\natexlab#1{#1}\fi

\bibitem[{{Aihara} {et~al.}(2018){Aihara}, {Arimoto}, {Armstrong}, {Arnouts},
  {Bahcall}, {Bickerton}, {Bosch}, \& et~al.}]{Aihara_etal18}
{Aihara}, H., {Arimoto}, N., {Armstrong}, R., {et~al.} 2018, \pasj, 70, S4

\bibitem[{{Amorisco} \& {Evans}(2012)}]{Amorisco_etal2012}
{Amorisco}, N.~C., \& {Evans}, N.~W. 2012, \apjl, 756, L2

\bibitem[{{Astropy Collaboration} {et~al.}(2013){Astropy Collaboration},
  {Robitaille}, {Tollerud}, {Greenfield}, {Droettboom}, {Bray}, {Aldcroft},
  {Davis}, {Ginsburg}, {Price-Whelan}, {Kerzendorf}, {Conley}, {Crighton},
  {Barbary}, {Muna}, {Ferguson}, {Grollier}, {Parikh}, {Nair}, {Unther},
  {Deil}, {Woillez}, {Conseil}, {Kramer}, {Turner}, {Singer}, {Fox}, {Weaver},
  {Zabalza}, {Edwards}, {Azalee Bostroem}, {Burke}, {Casey}, {Crawford},
  {Dencheva}, {Ely}, {Jenness}, {Labrie}, {Lim}, {Pierfederici}, {Pontzen},
  {Ptak}, {Refsdal}, {Servillat}, \& {Streicher}}]{Astropy}
{Astropy Collaboration}, {Robitaille}, T.~P., {Tollerud}, E.~J., {et~al.} 2013,
  \aap, 558, A33

\bibitem[{{Bate} {et~al.}(2015){Bate}, {McMonigal}, {Lewis}, {Irwin},
  {Gonzalez-Solares}, {Shanks}, \& {Metcalfe}}]{Bate_etal15}
{Bate}, N.~F., {McMonigal}, B., {Lewis}, G.~F., {et~al.} 2015, \mnras, 453, 690

\bibitem[{{Battaglia} {et~al.}(2006){Battaglia}, {Tolstoy}, {Helmi}, {Irwin},
  {Letarte}, {Jablonka}, {Hill}, {Venn}, {Shetrone}, {Arimoto}, {Primas},
  {Kaufer}, {Francois}, {Szeifert}, {Abel}, \& {Sadakane}}]{Battaglia_etal06}
{Battaglia}, G., {Tolstoy}, E., {Helmi}, A., {et~al.} 2006, \aap, 459, 423

\bibitem[{{Bersier}(2000)}]{Bersier_etal00}
{Bersier}, D. 2000, \apjl, 543, L23

\bibitem[{{Bertin} \& {Arnouts}(1996)}]{Bertin_etal96}
{Bertin}, E., \& {Arnouts}, S. 1996, \aaps, 117, 393

\bibitem[{{Bertin} {et~al.}(2002){Bertin}, {Mellier}, {Radovich}, {Missonnier},
  {Didelon}, \& {Morin}}]{Bertin_etal02}
{Bertin}, E., {Mellier}, Y., {Radovich}, M., {et~al.} 2002, in Astronomical
  Society of the Pacific Conference Series, Vol. 281, Astronomical Data
  Analysis Software and Systems XI, ed. D.~A. {Bohlender}, D.~{Durand}, \&
  T.~H. {Handley}, 228

\bibitem[{{Choi} {et~al.}(2016){Choi}, {Dotter}, {Conroy}, {Cantiello},
  {Paxton}, \& {Johnson}}]{Choi_etal16}
{Choi}, J., {Dotter}, A., {Conroy}, C., {et~al.} 2016, \apj, 823, 102

\bibitem[{{Coleman} {et~al.}(2004){Coleman}, {Da Costa}, {Bland-Hawthorn},
  {Mart{\'{\i}}nez-Delgado}, {Freeman}, \& {Malin}}]{Coleman_etal04}
{Coleman}, M., {Da Costa}, G.~S., {Bland-Hawthorn}, J., {et~al.} 2004, \aj,
  127, 832

\bibitem[{{Coleman} {et~al.}(2005){Coleman}, {Da Costa}, {Bland-Hawthorn}, \&
  {Freeman}}]{Coleman_etal05b}
{Coleman}, M.~G., {Da Costa}, G.~S., {Bland-Hawthorn}, J., \& {Freeman}, K.~C.
  2005, \aj, 129, 1443

\bibitem[{{Coleman} \& {de Jong}(2008)}]{Coleman_etal08}
{Coleman}, M.~G., \& {de Jong}, J.~T.~A. 2008, \apj, 685, 933

\bibitem[{{de Boer} {et~al.}(2013){de Boer}, {Tolstoy}, {Saha}, \&
  {Olszewski}}]{deBoer_etal13}
{de Boer}, T.~J.~L., {Tolstoy}, E., {Saha}, A., \& {Olszewski}, E.~W. 2013,
  \aap, 551, A103

\bibitem[{{de Boer} {et~al.}(2012){de Boer}, {Tolstoy}, {Hill}, {Saha},
  {Olszewski}, {Mateo}, {Starkenburg}, {Battaglia}, \&
  {Walker}}]{de_Boer_etal2012}
{de Boer}, T.~J.~L., {Tolstoy}, E., {Hill}, V., {et~al.} 2012, \aap, 544, A73

\bibitem[{{de Vaucouleurs} \& {Ables}(1968)}]{deVaucouleurs_etal68}
{de Vaucouleurs}, G., \& {Ables}, H.~D. 1968, \apj, 151, 105

\bibitem[{{del Pino} {et~al.}(2015){del Pino}, {Aparicio}, \&
  {Hidalgo}}]{del_Pino_etal2015}
{del Pino}, A., {Aparicio}, A., \& {Hidalgo}, S.~L. 2015, \mnras, 454, 3996

\bibitem[{{del Pino} {et~al.}(2017){del Pino}, {Aparicio}, {Hidalgo}, \&
  {{\L}okas}}]{del_Pino_etal2017}
{del Pino}, A., {Aparicio}, A., {Hidalgo}, S.~L., \& {{\L}okas}, E.~L. 2017,
  \mnras, 465, 3708

\bibitem[{{del Pino} {et~al.}(2013){del Pino}, {Hidalgo}, {Aparicio},
  {Gallart}, {Carrera}, {Monelli}, {Buonanno}, \&
  {Marconi}}]{del_Pino_etal2013}
{del Pino}, A., {Hidalgo}, S.~L., {Aparicio}, A., {et~al.} 2013, \mnras, 433,
  1505

\bibitem[{{DES Collaboration}(2018)}]{DES_2018}
{DES Collaboration}. 2018, in preparation

\bibitem[{{Dotter}(2016)}]{Dotter_etal16}
{Dotter}, A. 2016, \apjs, 222, 8

\bibitem[{{Drlica-Wagner} {et~al.}(2018){Drlica-Wagner}, {Sevilla-Noarbe},
  {Rykoff}, {Gruendl}, {Yanny}, {Tucker}, {Hoyle}, et~al., \& {DES
  Collaboration}}]{Drlica-Wagner_etal2018}
{Drlica-Wagner}, A., {Sevilla-Noarbe}, I., {Rykoff}, E.~S., {et~al.} 2018,
  \apjs, 235, 33

\bibitem[{{Flaugher} {et~al.}(2015){Flaugher}, {Diehl}, {Honscheid}, {Abbott},
  {Alvarez}, {Angstadt}, et~al., \& {DES Collaboration}}]{Flaugher_etal15}
{Flaugher}, B., {Diehl}, H.~T., {Honscheid}, K., {et~al.} 2015, \aj, 150, 150

\bibitem[{{Foreman-Mackey} {et~al.}(2013){Foreman-Mackey}, {Hogg}, {Lang}, \&
  {Goodman}}]{Foreman-Mackey_etal2013}
{Foreman-Mackey}, D., {Hogg}, D.~W., {Lang}, D., \& {Goodman}, J. 2013, \pasp,
  125, 306

\bibitem[{{Gaia Collaboration} {et~al.}(2016){Gaia Collaboration}, {Prusti},
  {de Bruijne}, {Brown}, {Vallenari}, {Babusiaux}, {Bailer-Jones}, {Bastian},
  {Biermann}, {Evans}, \& et~al.}]{GAIAmain}
{Gaia Collaboration}, {Prusti}, T., {de Bruijne}, J.~H.~J., {et~al.} 2016,
  \aap, 595, A1

\bibitem[{{Gallart} {et~al.}(2005){Gallart}, {Aparicio}, {Zinn}, {Buonanno},
  {Hardy}, \& {Marconi}}]{Gallart_etal05}
{Gallart}, C., {Aparicio}, A., {Zinn}, R., {et~al.} 2005, in IAU Colloq. 198:
  Near-fields cosmology with dwarf elliptical galaxies, ed. {H.~Jerjen \&
  B.~Binggeli}, 25--29

\bibitem[{{Goodman} \& {Weare}(2010)}]{Goodman_etal2010}
{Goodman}, J., \& {Weare}, J. 2010, Communications in Applied Mathematics and
  Computational Science, Vol.~5, No.~1, p.~65-80, 2010, 5, 65

\bibitem[{{Grillmair}(2009)}]{Grillmair_etal09}
{Grillmair}, C.~J. 2009, \apj, 693, 1118

\bibitem[{{Hodge}(1961{\natexlab{a}})}]{Hodge_etal61A}
{Hodge}, P.~W. 1961{\natexlab{a}}, \aj, 66, 83

\bibitem[{{Hodge}(1961{\natexlab{b}})}]{Hodge_etal61B}
---. 1961{\natexlab{b}}, \aj, 66, 249

\bibitem[{{Hodge} \& {Smith}(1974)}]{Hodge_etal74}
{Hodge}, P.~W., \& {Smith}, D.~W. 1974, \apj, 188, 19

\bibitem[{{Hunter}(2007)}]{Matplotlib}
{Hunter}, J.~D. 2007, Computing in Science and Engineering, 9, 90

\bibitem[{{Irwin} \& {Hatzidimitriou}(1995)}]{IH95}
{Irwin}, M., \& {Hatzidimitriou}, D. 1995, \mnras, 277, 1354

\bibitem[{{Kauffmann} {et~al.}(1993){Kauffmann}, {White}, \&
  {Guiderdoni}}]{Kauffmann_etal93}
{Kauffmann}, G., {White}, S.~D.~M., \& {Guiderdoni}, B. 1993, \mnras, 264, 201

\bibitem[{{King}(1962)}]{King_1962}
{King}, I. 1962, \aj, 67, 471

\bibitem[{{Letarte} {et~al.}(2006){Letarte}, {Hill}, {Jablonka}, {Tolstoy},
  {Fran{\c c}ois}, \& {Meylan}}]{Letarte_etal06}
{Letarte}, B., {Hill}, V., {Jablonka}, P., {et~al.} 2006, \aap, 453, 547

\bibitem[{{Letarte} {et~al.}(2010){Letarte}, {Hill}, {Tolstoy}, {Jablonka},
  {Shetrone}, {Venn}, {Spite}, {Irwin}, {Battaglia}, {Helmi}, {Primas},
  {Fran{\c c}ois}, {Kaufer}, {Szeifert}, {Arimoto}, \&
  {Sadakane}}]{Letarte_etal10}
{Letarte}, B., {Hill}, V., {Tolstoy}, E., {et~al.} 2010, \aap, 523, A17

\bibitem[{{{\L}okas}(2009)}]{Lokas_etal09}
{{\L}okas}, E.~L. 2009, \mnras, 394, L102

\bibitem[{{Martin} {et~al.}(2008){Martin}, {de Jong}, \&
  {Rix}}]{Martin_etal2008}
{Martin}, N.~F., {de Jong}, J.~T.~A., \& {Rix}, H.-W. 2008, \apj, 684, 1075

\bibitem[{{McConnachie}(2012)}]{McConnachie_12}
{McConnachie}, A.~W. 2012, \aj, 144, 4

\bibitem[{{Morganson} {et~al.}(2018){Morganson}, {Gruendl}, {Menanteau},
  {Carrasco Kind}, {Chen}, {Daues}, {Drlica-Wagner}, {Friedel}, {Gower},
  {Johnson}, {Johnson}, {Kessler}, {Paz-Chinch{\'o}n}, {Petravick}, {Pond},
  {Yanny}, {Allam}, {Armstrong}, {Barkhouse}, {Bechtol}, {Benoit-L{\'e}vy},
  {Bernstein}, {Bertin}, {Buckley-Geer}, {Covarrubias}, {Desai}, {Diehl},
  {Goldstein}, {Gruen}, {Li}, {Lin}, {Marriner}, {Mohr}, {Neilsen}, {Ngeow},
  {Paech}, {Rykoff}, {Sako}, {Sevilla-Noarbe}, {Sheldon}, {Sobreira}, {Tucker},
  \& {Wester}}]{Morganson_etal18}
{Morganson}, E., {Gruendl}, R.~A., {Menanteau}, F., {et~al.} 2018, ArXiv
  e-prints, arXiv:1801.03177

\bibitem[{{Nichols} {et~al.}(2012){Nichols}, {Lin}, \&
  {Bland-Hawthorn}}]{Nichols_etal12}
{Nichols}, M., {Lin}, D., \& {Bland-Hawthorn}, J. 2012, \apj, 748, 149

\bibitem[{{Olszewski} {et~al.}(2006){Olszewski}, {Mateo}, {Harris}, {Walker},
  {Coleman}, \& {Da Costa}}]{Olszewski_etal06}
{Olszewski}, E.~W., {Mateo}, M., {Harris}, J., {et~al.} 2006, \aj, 131, 912

\bibitem[{{Pasetto} {et~al.}(2011){Pasetto}, {Grebel}, {Berczik}, {Chiosi}, \&
  {Spurzem}}]{Pasetto_etal11}
{Pasetto}, S., {Grebel}, E.~K., {Berczik}, P., {Chiosi}, C., \& {Spurzem}, R.
  2011, \aap, 525, A99

\bibitem[{{Pietrzy{\'n}ski} {et~al.}(2009){Pietrzy{\'n}ski}, {G{\'o}rski},
  {Gieren}, {Ivanov}, {Bresolin}, \& {Kudritzki}}]{Pietrzynski_etal09}
{Pietrzy{\'n}ski}, G., {G{\'o}rski}, M., {Gieren}, W., {et~al.} 2009, \aj, 138,
  459

\bibitem[{{Plummer}(1911)}]{Plummer_1911}
{Plummer}, H.~C. 1911, \mnras, 71, 460

\bibitem[{{Pont} {et~al.}(2004){Pont}, {Zinn}, {Gallart}, {Hardy}, \&
  {Winnick}}]{Pont_etal04}
{Pont}, F., {Zinn}, R., {Gallart}, C., {Hardy}, E., \& {Winnick}, R. 2004, \aj,
  127, 840

\bibitem[{{Rizzi} {et~al.}(2007){Rizzi}, {Held}, {Saviane}, {Tully}, \&
  {Gullieuszik}}]{Rizzi_etal07}
{Rizzi}, L., {Held}, E.~V., {Saviane}, I., {Tully}, R.~B., \& {Gullieuszik}, M.
  2007, \mnras, 380, 1255

\bibitem[{{Rockosi} {et~al.}(2002){Rockosi}, {Odenkirchen}, {Grebel}, {Dehnen},
  {Cudworth}, {Gunn}, {York}, {Brinkmann}, {Hennessy}, \&
  {Ivezi{\'c}}}]{Rockosi_etal02}
{Rockosi}, C.~M., {Odenkirchen}, M., {Grebel}, E.~K., {et~al.} 2002, \aj, 124,
  349

\bibitem[{{Schlegel} {et~al.}(1998){Schlegel}, {Finkbeiner}, \&
  {Davis}}]{Schlegel_etal1998}
{Schlegel}, D.~J., {Finkbeiner}, D.~P., \& {Davis}, M. 1998, \apj, 500, 525

\bibitem[{{Sersic}(1968)}]{Sersic_1968}
{Sersic}, J.~L. 1968, {Atlas de Galaxias Australes}

\bibitem[{{Shapley}(1938)}]{Shapley_etal38}
{Shapley}, H. 1938, \nat, 142, 715

\bibitem[{{Sheldon}(2014)}]{Sheldon_14}
{Sheldon}, E.~S. 2014, \mnras, 444, L25

\bibitem[{{Shipp} {et~al.}(2018){Shipp}, {Drlica-Wagner}, {Balbinot},
  {Ferguson}, {Erkal}, {Li}, {Bechtol}, {Belokurov}, {Buncher}, {Carollo},
  {Carrasco Kind}, et~al., \& {the DES Collaboration}}]{Shipp_etal18}
{Shipp}, N., {Drlica-Wagner}, A., {Balbinot}, E., {et~al.} 2018, ArXiv
  e-prints, arXiv:1801.03097

\bibitem[{{Stetson}(1987)}]{Stetson_1987}
{Stetson}, P.~B. 1987, \pasp, 99, 191

\bibitem[{{Van Der Walt} {et~al.}(2011){Van Der Walt}, {Colbert}, \&
  {Varoquaux}}]{NumPy}
{Van Der Walt}, S., {Colbert}, S.~C., \& {Varoquaux}, G. 2011, ArXiv e-prints,
  arXiv:1102.1523

\bibitem[{{Walker} {et~al.}(2006){Walker}, {Mateo}, {Olszewski}, {Bernstein},
  {Wang}, \& {Woodroofe}}]{Walker_etal06}
{Walker}, M.~G., {Mateo}, M., {Olszewski}, E.~W., {et~al.} 2006, \aj, 131, 2114

\bibitem[{{Wang} {et~al.}(2017){Wang}, {Fattahi}, {Cooper}, {Sawala},
  {Strigari}, {Frenk}, {Navarro}, {Oman}, \& {Schaller}}]{Wang_etal17}
{Wang}, M.-Y., {Fattahi}, A., {Cooper}, A.~P., {et~al.} 2017, \mnras, 468, 4887

\bibitem[{{Yozin} \& {Bekki}(2012{\natexlab{a}})}]{Yozin_etal12}
{Yozin}, C., \& {Bekki}, K. 2012{\natexlab{a}}, \apjl, 756, L18

\bibitem[{{Yozin} \& {Bekki}(2012{\natexlab{b}})}]{Yozin_etal2012}
---. 2012{\natexlab{b}}, \apjl, 756, L18

\bibitem[{{Yuan} {et~al.}(2016){Yuan}, {Qian}, \& {Jing}}]{Yuan_etal16}
{Yuan}, Z., {Qian}, Y.-Z., \& {Jing}, Y.~P. 2016, \mnras, 456, 3253

\end{thebibliography}

\end{document}